\documentclass[manuscript,screen]{acmart}

\renewcommand\footnotetextcopyrightpermission[1]{} 
\usepackage{amsmath,amsfonts}
\usepackage{algorithmic}
\usepackage{graphicx}
\usepackage{todonotes}
\usepackage{textcomp}
\usepackage{xcolor}
\usepackage{booktabs}
\usepackage{caption}
\usepackage{subcaption}
\usepackage{array}
\newcommand{\fuseconv}{FuSeConv}
\newcommand{\rtos}{ST-OS}
\newcommand{\stos}{ST-OS}
\newcommand{\nos}{NOS}
\newcommand{\dwsepconv}{depthwise-separable convolution}
\newcommand{\dwsepconvs}{depthwise-separable convolutions}
\newcommand{\fusehalf}{FuSe-Half}
\newcommand{\fusefull}{FuSe-Full}

\pdfoutput=1 

\AtBeginDocument{%
  \providecommand\BibTeX{{%
    \normalfont B\kern-0.5em{\scshape i\kern-0.25em b}\kern-0.8em\TeX}}}

\begin{document}

\title{Design and Scaffolded Training of an Efficient DNN Operator for Computer Vision on the Edge}


\author{Vinod Ganesan}
\email{vinodg@cse.iitm.ac.in}
\affiliation{%
  \institution{Indian Institute of Technology, Madras}
  \country{India}
}
\author{Pratyush Kumar}
\email{pratyush@cse.iitm.ac.in}
\affiliation{%
  \institution{Microsoft Research and Indian Institute of Technology, Madras}
  \country{India}
}

%
%
%
%
%
%
%

\renewcommand{\shortauthors}{Ganesan, et al,.}

\begin{abstract}
Massively parallel systolic arrays and resource-efficient depthwise separable convolutions are two promising hardware and software techniques to accelerate DNN inference on the edge. 
Interestingly, their combination is inefficient: Computational patterns of depthwise separable convolutions do not exhibit a rhythmic systolic flow and lack sufficient data reuse to saturate systolic arrays. 
In this paper, we formally analyse this inefficiency and propose an efficient operator, an optimal hardware dataflow, and a superior training methodology towards alleviating this. 
The efficient operator, called Fully-Separable Convolutions (\fuseconv) \footnote{First introduced in \cite{selvam2021fuseconv}}, is a drop-in replacement for \dwsepconvs. 
\fuseconv~generalizes factorization of convolution fully along their spatial and depth dimensions. 
The resultant computation is systolic and efficiently maps to systolic arrays. 
The optimal hardware dataflow, called \emph{Spatial-Tiled Output Stationary} (\stos), maximizes the efficiency of \fuseconv~on systolic arrays.
It maps independent convolutions to rows of the systolic array to maximize resource-utilization with negligible VLSI overheads.
\emph{Neural Operator Scaffolding (NOS)} scaffolds the training of \fuseconv~operators by distilling knowledge from the more expensive depthwise separable convolution operation. 
This bridges the accuracy gap between \fuseconv~networks and networks with \dwsepconvs.
Additionally, \nos~can be combined with Neural Architecture Search (NAS) to trade-off latency and accuracy.

The hardware-software co-design of \fuseconv~with \stos~achieves a significant speedup of $4.1-9.25\times$ with state-of-the-art efficient networks for the ImageNet dataset. 
The parameter efficiency of \fuseconv~and its significant out-performance over \dwsepconvs~on systolic arrays illustrates their promise as a strong solution on the edge. 
Training \fuseconv~networks with \nos~achieves accuracy comparable to the \dwsepconv~baselines. 
Further, by combining \nos~with NAS, we design networks that define state-of-the-art models improving on both accuracy and latency for computer vision on systolic arrays. 

\end{abstract}

%


\maketitle

\section{Introduction}
Deep Neural Networks (DNNs) continue to advance the field of Machine Learning and are being pervasively deployed.
This widespread use comes at the cost of high computational demands which at times exceeds available hardware resources. 
This is particularly a concern on the edge with devices built with strict area, cost, and power constraints. 
Several classes of methods have been explored to address this challenge of efficient inference on the edge, as summarised in this work \cite{xu2018scaling}.
These include model compression \cite{han2015deep,he2018amc}, efficient operator design \cite{howard2017mobilenets,sandler2018mobilenetv2,iandola2016squeezenet,gholami2018squeezenext}, hardware extensions such as leveraging sparsity \cite{sen2018sparce,ganesan2020sparsity,han2016eie}, and domain-specific hardware accelerators \cite{jouppi2017datacenter,jouppi2020domain,chung2018serving}. 
Amongst these, efficient operators and domain-specific hardware accelerators have been particularly successful. 

An example of an efficient operator is the \textit{depthwise separable convolution} for computer vision. 
In this operator, standard convolution is factorized into independent convolutions on each input channel (called depthwise convolution) followed by a 1x1 pointwise convolution.
This factorization significantly reduces the number of parameters and operations as with the MobileNet \cite{howard2017mobilenets, sandler2018mobilenetv2, howard2019searching} family of networks: MobileNet-V3 consumes $14.5\times$ fewer MACs but delivers 1.6\% higher accuracy than DenseNet-121 \cite{huang2017densely} on ImageNet.
An example of a domain-specific hardware accelerator is the use of \textit{systolic arrays} \cite{kung1980systolic}.
The inclusion of systolic arrays is now a common-place design pattern to accelerate kernels such as matrix multiplication and convolution using a grid of multiply-accumulate units (MACs).
Its efficiency is exemplified by TPUs \cite{jouppi2017datacenter}: At the time of release, the TPUv1 provided 25 to 29 times higher performance-per-watt than comparable GPUs \cite{jouppi2017datacenter}.

Surprisingly however, the combination of these two efficient ideas, (i.e.) depthwise convolution running on systolic arrays, is inefficient. 
For instance, MobileNet-V2 runs only 1.3x faster than ResNet-50 on a $32 \times 32$ systolic array despite having 12$\times$ fewer MAC operations. 
This incommensurate scaling has been identified earlier on EdgeTPU running EfficientNet \cite{gupta2020accelerator} and on SqueezeNext \cite{gholami2018squeezenext}. For instance, the authors of \cite{gupta2020accelerator} report cases where a spatial convolution with significantly more trainable parameters and MAC operations performs $1.46\times$ better than a depthwise convolution due to better utilization of the EdgeTPU hardware. 
In this work, we study this incommensurate scaling and pose three questions: (1) why are depthwise separable convolutions inefficient on systolic arrays?, (2) is there a HW/SW co-design approach to address this inefficiency?, and (3) can we adapt model training to better fit this new design?

For the first question, we discuss the formalism of systolic algorithms  \cite{song1994systolic}, a subclass of Regular Iterative Algorithms \cite{rao1988regular} that efficiently run on systolic architectures. 
We show that a \textit{depthwise separable convolution} is not a systolic algorithm.
Consequently, mapping it on a systolic array requires im2col \cite{chellapilla2006high} transformation to convert depthwise convolutions to matrix multiplications. 
However, this conversion replicates a large number of values, and unlike 2D convolution does not have sufficient data-reuse to saturate the systolic array. 

Having identified the source of inefficiency, we propose a HW/SW co-design to address that consists of two parts: (a) an efficient operator - Fully-Separable Convolution (\fuseconv), and (b) a modified hardware dataflow -  Spatial-Tiled Output Stationary (\stos). 
\fuseconv~is a drop-in replacement for depthwise separable convolutions.
Separable convolutions factorize one of the four dimensions of a convolution filter. 
In \fuseconv, we fully generalize this factorization to 1D convolutions across both spatial (horizontal and vertical) and channel (width) dimensions. 
This is efficient as 1D convolutions are systolic algorithms and do not require im2col transformations.
\stos~is a novel hardware dataflow that exploits the parallelism offered by \fuseconv~by mapping each 1D convolution on to one row of the systolic array.
This is enabled by adding a weight-broadcast link to each row of the systolic array.
The hardware cost for supporting \stos~is small: a $16\times 16$ sized systolic array incurs only $3.2\%$ area and $6.7\%$ power overheads when synthesized on a 22nm node.

To train more accurate models with the proposed \fuseconv~operator, we propose \emph{Neural Operator Scaffolding} (\nos). 
The key insight is to distill knowledge from the more expensive depthwise filters to FuSe filters, progressively during the training phase.
We achieve this with a scaffolded network containing adapter matrices between depthwise and FuSe filters.
Once the scaffolded training is complete, we can either retain only the FuSe filters or a combination of depthwise and FuSe filters. 
We search for these combinations with evolutionary search \cite{real2017large} and Neural Architecture Search (NAS) \cite{cai2019once} to trade-off model accuracy and latency.

We evaluate \fuseconv~and \stos~on a large set of popular efficient networks and metrics. 
To obtain accurate latency measures, we extend the systolic array simulator SCALE-Sim \cite{samajdar2018scale} to support the \stos~dataflow.
For existing efficient networks such as MobileNet (V1, 2 ,3) and MnasNet, replacing all \dwsepconvs~ with \fuseconv~variants leads to substantial drops of 4.15-9.25 times in inference time on a $16\times 16$ systolic array.
\fuseconv~networks trained with \nos~on the ImageNet dataset are up to $1.5-2\%$ more accurate than conventional training and are as accurate as their baseline \dwsepconv~networks.  
When \nos~is combined with evolutionary search \cite{real2017large} and NAS \cite{cai2019once}, we train networks within a superior accuracy and efficiency trade-off space. 
For instance, \nos~coupled with once-for-all \cite{cai2019once}, a state-of-the-art NAS approach, discovers a network that has $77.2\%$ accuracy on ImageNet and $3.82$ ms latency on a $16\times 16$ systolic array, improving on both metrics in comparison to both hand-crafted and automatically designed networks.
In summary, we establish that the combination of \fuseconv~operator, \stos~dataflow, and \nos~training method significantly improves the efficiency of computer vision models on systolic arrays. 


The key contributions in this paper are as follows: 
\begin{itemize}
    \item 
    We identify the problem of incommensurate scaling with two efficient methods, (i.e.) systolic arrays executing depthwise separable convolutions.
    We show that depthwise convolution is not a systolic algorithm and its mapping requires transformations that cannot utilize two-dimensional systolic arrays. 
    \item We propose \emph{Fully-Separable Convolutions} (\fuseconv) as a drop-in replacement operator for \dwsepconvs, and \emph{Spatial-Tiled Output Stationary} (\stos) as a custom dataflow to improve the efficiency of \fuseconv~on systolic arrays.  
    \item We propose \emph{Neural Operator Scaffolding} (\nos), to progressively train cheap \fuseconv~operators by distilling information from costly depthwise convolution operators. 
    \item We extend the open-source SCALE-Sim simulator \cite{samajdar2018scale} to support \dwsepconv, \fuseconv, and \stos.
    \item We show that the VLSI overheads of adding the \stos~dataflow are minimal: a $16\times 16$ sized systolic array incurs only $3.2\%$ area and $6.7\%$ power overheads when synthesized on a 22nm node.
    \item On state-of-the-art efficient models, we show that replacing \dwsepconv~ with \fuseconv~ leads to a significant $4.1-9.25\times$ improvements in inference time on a $16 \times 16$ systolic array.
    \item With \nos, the accuracy of \fuseconv~networks is improved by up to $1.5-2\%$ and this bridges the gap to baseline networks that use more expensive \dwsepconvs. 
    \item With \nos~coupled with evolutionary search and NAS, we find models that improve on both accuracy and latency, creating the state-of-the-art computer vision models for systolic arrays. 
\end{itemize}

The rest of the paper is organized as follows. 
In Section \ref{sec:sysalgo}, we formally analyse the inefficiency of \dwsepconvs~on systolic arrays. 
In Sections \ref{sec:fuseconv_stos} and \ref{sec:nos}, we discuss our core contribution of the HW/SW co-design solution and \nos, respectively.
In Section \ref{sec:experimental}, we detail our experimental methodology used to evaluate our proposal and present the results in Section \ref{sec:evaluation}.
We discuss the related work in Section \ref{sec:related} and conclude in Section \ref{sec:conclusion}.
\begin{figure*}
    \centering

    \includegraphics[width=0.9\textwidth]{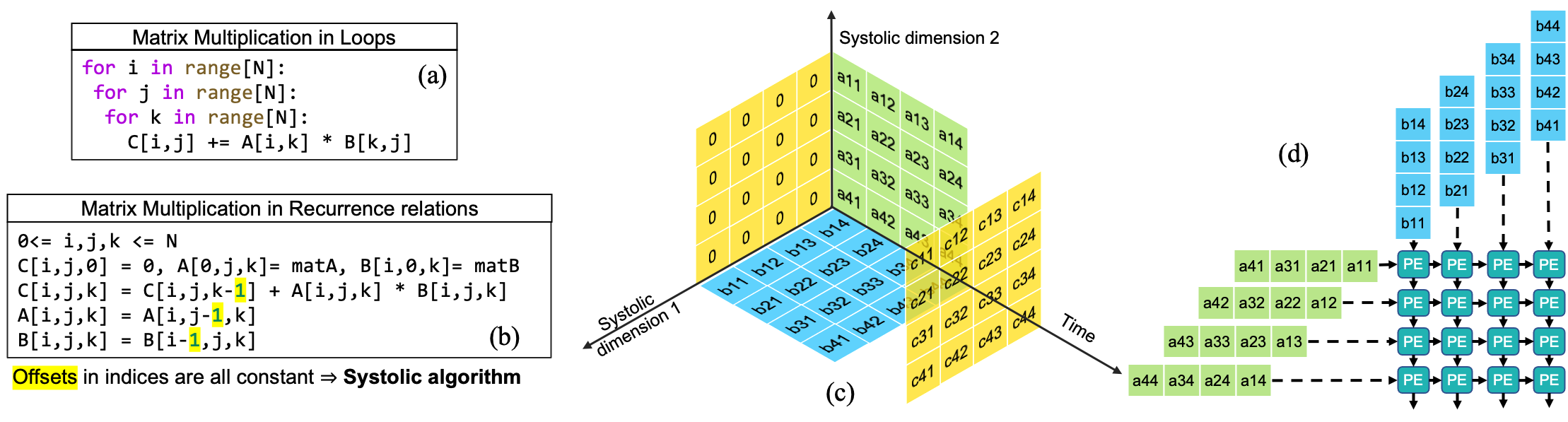}
    \caption{(a). Loop representation of matrix multiplication; (b). Corresponding recurrence relations; (c). Geometric representation of the recurrence relation; (d). Output-stationary dataflow mapping to a systolic array}
    \label{fig:mm}
  
\end{figure*}

\label{sec:introduction}


\section{Inefficiency of depthwise separable convolutions on systolic arrays}
\label{sec:sysalgo}
In this section, we answer the first research question and show why \dwsepconv~has poor scaling performance on systolic arrays.
We also explain this finding in contrast to the widespread use of systolic arrays for standard convolution.
We first begin with a background on these topics.


\subsection{Background - Systolic Arrays and Algorithms}

\noindent \textbf{Systolic Arrays} A systolic array \cite{kung1982systolic} is a parallel computing hardware, constituted of a large number of homogeneous Processing Elements (PEs). 
These PEs are interconnected in a regular arrangement, such as a rectangular grid, facilitating local communication with neighbours.  
The word `systolic' implies rhythmic patterns in communication and computation, which are globally synchronous.
This regular arrangement constraints how data flows amongst PEs thereby restricting the number of applications that can be mapped onto systolic arrays. 
However, embarrassingly parallel and regular applications such as matrix multiplication can be efficiently mapped onto systolic arrays.
The main benefits of systolic arrays come from the nearest neighbour communication leading to low main memory dependence, and low control overhead. 
Currently, systolic arrays are quite common design patterns in accelerators \cite{jouppi2017datacenter, XlinixDNNProcessor:online}.

\noindent \textbf{Regular Iterative Algorithms}
Single-Assignment Language (SAL) \cite{gross1987mapping} is a programming style requiring that no variable is assigned a value more than once (i.e. c = c - b is not allowed). 
Regular Iterative algorithms (RIA) are a subclass of SALs that map to architectures with a regular array of processors, each performing the same computation and exhibiting simple inter-processor communication \cite{rao1988regular}. 
Systolic algorithms are a further subclass of RIAs such that they can be synthesized to a systolic array \cite{wan1996systolic}.
We discuss systolic algorithms with the example of matrix multiplication, as studied in the pioneering work of Kung and Leiserson \cite{kung1980systolic}.
In Fig.~\ref{fig:mm}(a) we show the standard for-loop implementation of a matrix-matrix multiplication. 
This is transformed into recurrence relations as shown in Fig.~\ref{fig:mm}(b).
These relations are said to define an RIA, since they satisfy three conditions: 
(a) each variable is defined by a name and a set of indices (in this case 3), 
(b) each variable is assigned a value just once (SAL \cite{gross1987mapping}). and
(c) for each recurrence relation, the difference between the indices of the variable in the LHS and each variable in the RHS is a constant.
For instance, in the equation for $C[i, j, k]$, the differences in indices (called index offsets) to the three variables $A$, $B$, and $C$ in the RHS are the constant vectors $[0, 0, 0], [0, 0, 0], $ and $[0, 0, -1]$, respectively.

\noindent \textbf{Mapping algorithms into systolic arrays}
The computation of the recurrence relations can be visualized as shown in Fig.~\ref{fig:mm}(c).
Each point in the 3D space corresponds to a combination of the indices $i, j, k$ starting at the origin $(0,0,0)$.
Inputs $A$ and $B$ are initialized in respective planes while output $C$ is initialized with 0s. 
All values are propagated through to other points as per the recurrence relations. 
The computations for updating the values are mapped on to three dimensions.
The dimensions $i$ and $j$, along which $A$ and $B$ are propagated, are systolic dimensions along which PEs are arranged.
The dimension $k$ along which $C$ is updated is marked as the `time' dimension, ending with the final computed value as shown. 
Assigning such dimensions is equivalent to \textit{mapping} the algorithm on to a 2D systolic array as shown in Fig.~\ref{fig:mm}(d).
In the mapping, matrices $A$ and $B$ are input along rows and columns (the two systolic dimensions), respectively, while output $C$ is computed in each processing element (PE) over time (the time dimension).
Since the output remains stationary in the PEs, this dataflow is referred to as \textit{output stationary}. 
We can similarly visualise other popular dataflows such as input and weight stationary. 

\noindent \textbf{Spatial and Depthwise Separable Convolutions}
In \textit{spatial convolution}, an input of size $W\times H\times C$ is convolved with a filter of size $K\times K\times C$ to obtain an output of size $N\times M\times 1$, where $N=W-K+1$ and $M=H-K+1$. 
The output of $C'$ convolution filters are stacked to obtain an output of size $N\times M\times C'$.
Depthwise-separable convolution decomposes the $K\times K\times C$ spatial convolution into two parts: (a) $K\times K\times C$ \textit{depthwise convolution} and (b) $1\times 1$ pointwise convolution.
In \textit{depthwise convolution}, an input of size $W\times H\times C$ is convolved with a filter of size $K\times K\times C$ channel-wise, \textit{i.e.}, every $W\times H$ channel is convolved with the respective $K\times K$ channel in the filter to obtain an output of size $N\times M\times C$.
This is followed by a \textit{point-wise convolution} with $C'$ filters of size $1\times 1\times C$ to obtain an output of size $N\times M\times C'$.
For both these illustrated cases, the input and output sizes are the same.
However, there is a major difference in the number of operations: Standard convolution has a total of $NMC'K^2C$ operations, while depthwise separable convolution has $NMC\times (K^2+C')$ operations.
This reduction in number of operations with depthwise separable convolution translates to reduced inference time at comparable accuracy values \cite{howard2017mobilenets,sandler2018mobilenetv2}.

\subsection{2D convolutions are not systolic algorithms}

\begin{figure}
    \centering
    \includegraphics[width=0.7\textwidth]{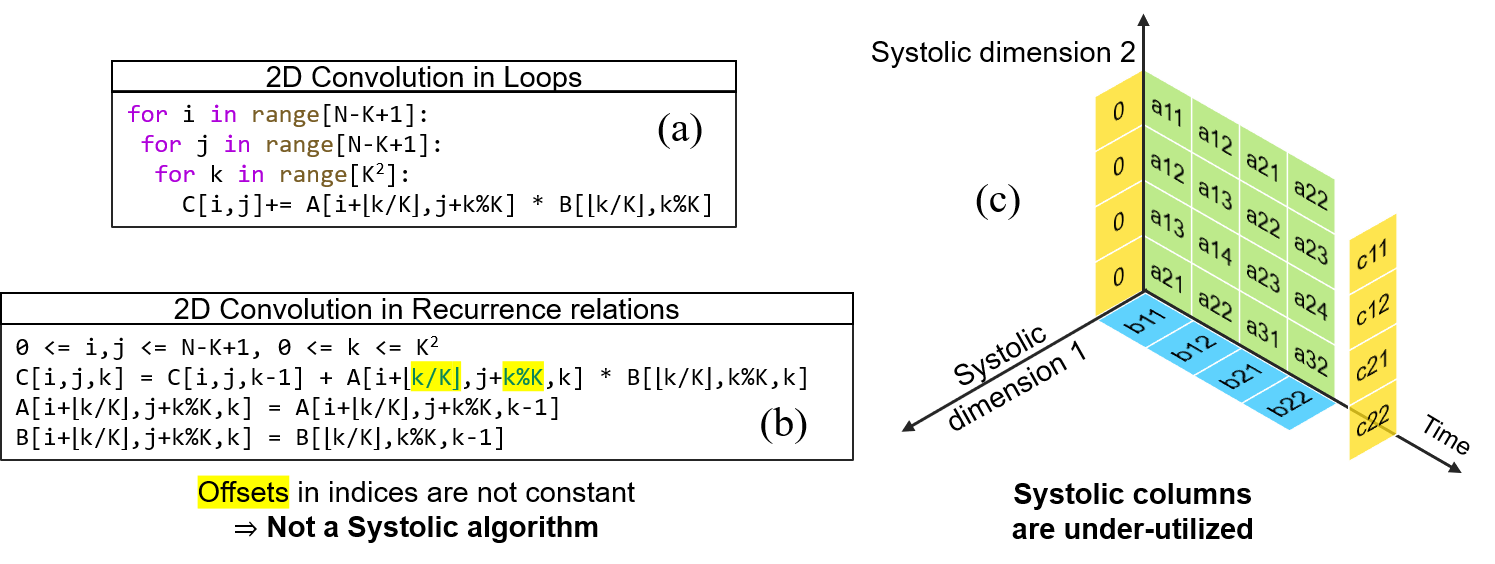}
    \caption{(a) Loop representation of 2D convolution; (b) Corresponding recurrence relations; (c) Geometric representation after transforming 2D convolution to a systolic algorithm.}
    \label{fig:convsysalgo}
\end{figure}

As described earlier, depthwise convolution requires independent convolutions between 2D slices of the input with 2D kernels, henceforth referred to as a 2D convolution. 
2D convolution can be written in loops as shown in Fig.~\ref{fig:convsysalgo}(a): $A$ is the input feature map, $B$ the weight kernel, and $C$ the output.
Is this a systolic algorithm?
We first attempt to transform it into a Regular Iterative Algorithm (RIA) which as a set contains all systolic algorithms.
Fig.~\ref{fig:convsysalgo}(b) shows the recurrence relations for 2D convolution. 
Like in the case of matrix multiplication, we have added a third index to satisfy the single assignment property.
Unlike the case of matrix multiplication, however, we observe that the index offsets between LHS and RHS are not constants.
For instance, in the recurrence relation for $C$, the index offset to $A$ is given as $[\lfloor k/K \rfloor, k\%K, 0]$.
Since this index offset depends on the index $k$, it violates the important requirement for a RIA.

Can this specification be refactored in some other way to satisfy RIA's requirement?
Note that computing the output at index $(i, j)$ requires summing up $K^2$ products.
We can map these computations to $K^2$ values of the $k$ index.
Computing a sum of these products implies a single offset dependence between $C[i, j, k]$ and $C[i, j, k+1]$.
For matrices $A$ and $B$, the computation of these products requires input across a grid of $K \times K$ values.
Independent of the order in which these values are accessed, their $i, j$ indices will depend on the index $k$.
However, all these products are summed to the same $i, j$ index of $C$.
Thus, in the same recurrence relation, the $i, j$ index of $C$ remain constant while those of $A, B$ depend on $k$, violating the criterion for constant index offsets.
Thus, 2D convolution cannot be written as an RIA, and consequently depthwise convolution is not a systolic algorithm.

Though 2D convolution is not a systolic algorithm, standard convolution with multiple channels can be efficiently mapped on to systolic arrays with one of two tricks.
The first approach is to transform the matrices with im2col and the second is to use channel-wise operations. 
However, neither approach applies to depthwise convolution, as we discuss in the following two subsections. 

\subsection{Transformation im2col and Data Reuse}
\begin{figure}
    \centering
    \includegraphics[width=0.5\textwidth]{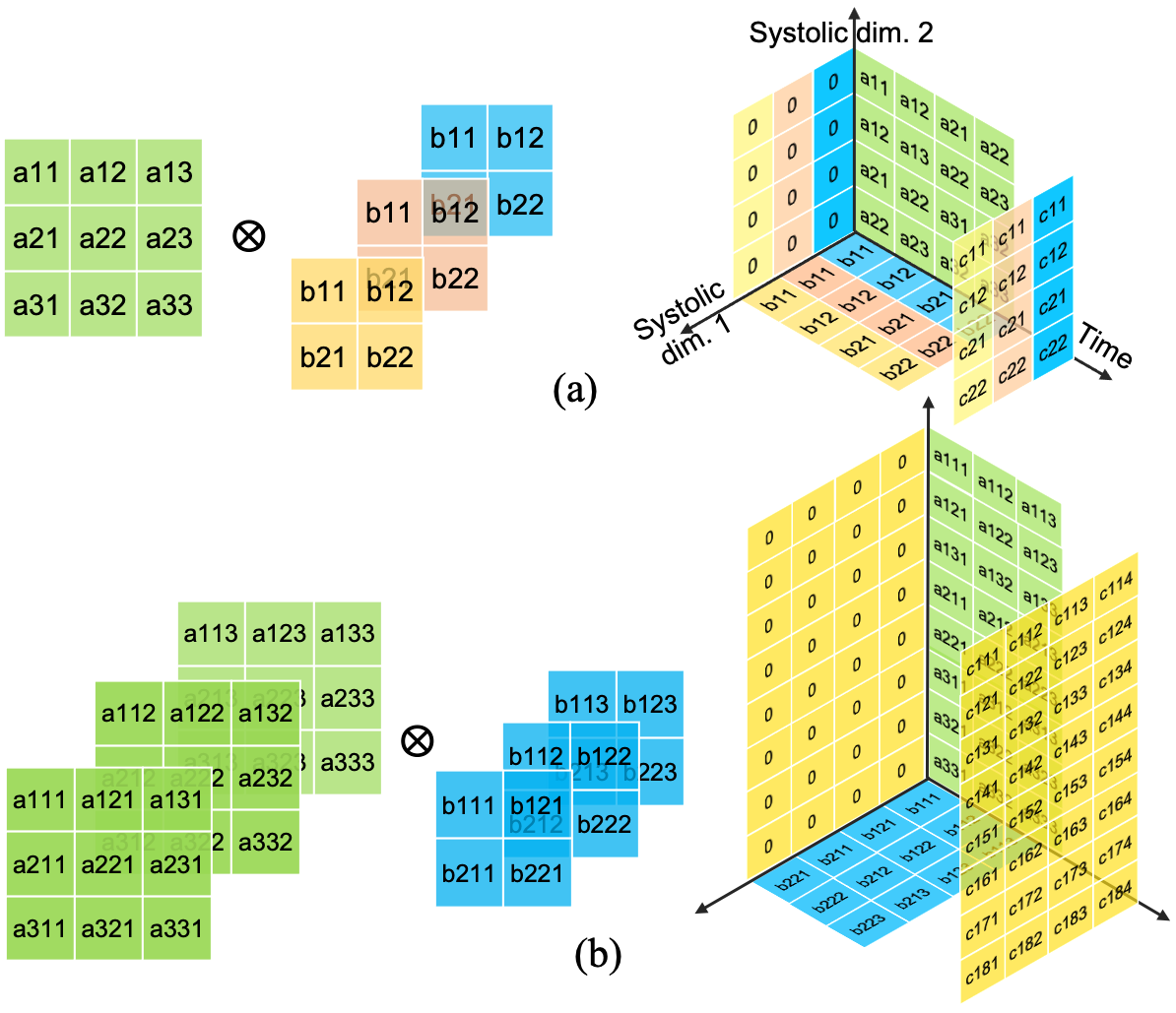}
    \caption{Two methods that enable execution of standard convolution on systolic arrays: (a) Input reuse across filters, (b) Ordering operations along channels.}
    \label{fig:channelwise}
\end{figure}

Consider a transformation of $A$ such that each set of $K\times K$ values required in each step of convolution is stored in a row.
With this transformation, the index offsets between $C$ and $A$ become constant.
This is the approach with im2col \cite{jia2014caffe} which creates a larger matrix $A'$ from $A$ with repeating entries and a flattened $B$ matrix.
The 2D convolution operation on these modified matrices is a systolic algorithm as shown in Fig.~\ref{fig:convsysalgo}(c).
Note that this computation does not scale on systolic dimension 1, {i.e.}, when mapped to a 2D systolic array it would only use a single column. 
With standard convolution with multiple channels and filters, this is not a problem as the same input channel has to be convolved with weight matrices from multiple filters utilizing all columns and benefiting from data reuse. 
This is shown in Fig.~\ref{fig:channelwise}(a), wherein the filters scale along systolic dimension 1 achieving high utilization.
However, this trick does not apply to depthwise convolution, wherein channel-wise decomposition which was designed for parameter efficiency does not enable data reuse or column utilization.
Thus, with depthwise convolution only a single column of a systolic array can be utilized leading to significantly poor performance.



\subsection{Channel-Wise Operations}
To avoid the expensive im2col transformation, an alternative approach maps standard convolution into channel-wise operations on systolic arrays \cite{jouppi2017datacenter}.
Specifically, standard convolution is implemented as a dot product of vectors of size $C$ along channels from the input and filter matrices.
The corresponding mapping on to a systolic array is shown in Fig.~\ref{fig:channelwise}(b).
The generated output needs to be reduced with an adder tree (usually a part of most systolic array accelerators) to obtain the final output.
For instance, the TPUv1 contains an adder tree in the periphery to reduce the incoming partial sums \cite{jouppi2017datacenter}.
In this case too, we are able to utilize both systolic dimensions.
Again this trick is not applicable to depthwise convolution. 
By definition, the convolution proceeds one channel at a time, precluding any channel-wise operations. \\

In summary, we discussed the formalism of RIA and how it applies to efficient utilization of systolic arrays. 
We showed that the 2D convolution is not a systolic algorithm.
However, the standard convolution on multiple channels and filters can still be mapped to systolic arrays due to the two tricks of im2col transformation or channel-wise operations. 
We showed that neither of these tricks is applicable to depthwise convolution due to lack of data-reuse and no channel-wise operations. 
Consequently, when mapping depthwise convolution to a systolic array, only a single systolic dimension can be utilized. 
This formally explains the poor performance of networks such as MobileNet on EdgeTPUs \cite{gupta2020accelerator} and motivates the design of operators which are simultaneously parameter-efficient and fully utilize two-dimensional systolic arrays. 


\section{HW/SW co-design: \fuseconv~and \stos}
\label{sec:fuseconv_stos}
\noindent 

In this section, we answer our second research question on addressing the inefficiency established in section \ref{sec:sysalgo}. 
We introduce a HW/SW co-design solution comprising of the \fuseconv~operator, its different variants, and its mapping to the systolic array architecture using the \stos~dataflow.

\subsection{The \fuseconv~operator}

As we discussed in the last section, \dwsepconv~have fewer parameters and require fewer MACs by decomposing spatial convolutions. 
However, the resultant computation is not systolic and does not benefit from either filter reuse or channel-wise mapping. 
We design \fuseconv~with the goal of deriving the benefits of decomposition from \dwsepconvs~while fully utilizing the systolic array.

Figure \ref{fig:fuseconv_stos} (a) shows the \fuseconv~operator in relation to \dwsepconv. 
A \dwsepconv~decomposes a $K\times K\times C\times C'$ spatial convolution along the depth axis by first performing $C$ $K\times K$ independent 2D convolutions (depthwise), followed by $C'$ $1\times 1\times C$ pointwise convolutions for channel-wise aggregation. 
In \fuseconv, we further factorize these independent 2D depthwise convolutions along the $K\times K$ spatial axes into $K\times 1$ row filters and $1\times K$ column filters.
At this point, we have two possible choices or variants, drawing inspiration from the work on grouped convolutions \cite{krizhevsky2012imagenet}. 
One, we can let row and column filters operate independently on the $C$ input channels creating an output feature map with $2C$ channels. 
Two, we restrict row filters to operate on $C/2$ input channels, and column filters to operate on the other $C/2$ channels, thereby creating $C$ output channels. 
We call the former the \fusefull~variant and the latter the \fusehalf~variant. 
\fusefull~clearly has more computation and parameters and would likely have higher accuracy. 
\fusehalf~is parameter efficient and retains the output tensor dimension with $C$ channels.
The resultant output, in either case, is passed through $C'$ pointwise filters to get the final output feature map.
Given our aim of efficient inference on the edge, we focus our analysis on the parameter-efficient \fusehalf~variant, but compare it against the more expensive \fusefull~variant.
Thus, when specifying \fuseconv~we by default refer to the \fusehalf~variant.

In summary, we propose \fuseconv~to decompose standard convolution fully - into channels and each of the two spatial dimensions. 
The operator is a drop-in replacement for \dwsepconv~retaining the same input and output channels, and thus can be used in mobile bottleneck layers \cite{sandler2018mobilenetv2,howard2019searching}.
In the following, we argue why \fuseconv~is an efficient operator for systolic arrays. 

\subsection{Efficiency of \fuseconv}

\subsubsection{Fewer Parameters}
The total number of parameters (both depthwise and pointwise) in \dwsepconv~are $C \times (K^2 + C')$.
For \fusehalf~this number reduces to $C \times (K + C')$.
Similarly, the number of operations reduces from $NMC \times (K^2 + C')$ to $NMC \times (K + C')$.
Thus, both model size and inference time is expected to reduce with \fusehalf.
Note that these gains are on top of depthwise convolution that already reduces the number of parameters and operations in comparison with standard convolution and is used in state-of-the-art networks for mobile devices. 

\begin{figure}[t!]
    \centering
    \begin{subfigure}[t]{0.5\textwidth}
        \centering
        \includegraphics[width=0.9\textwidth]{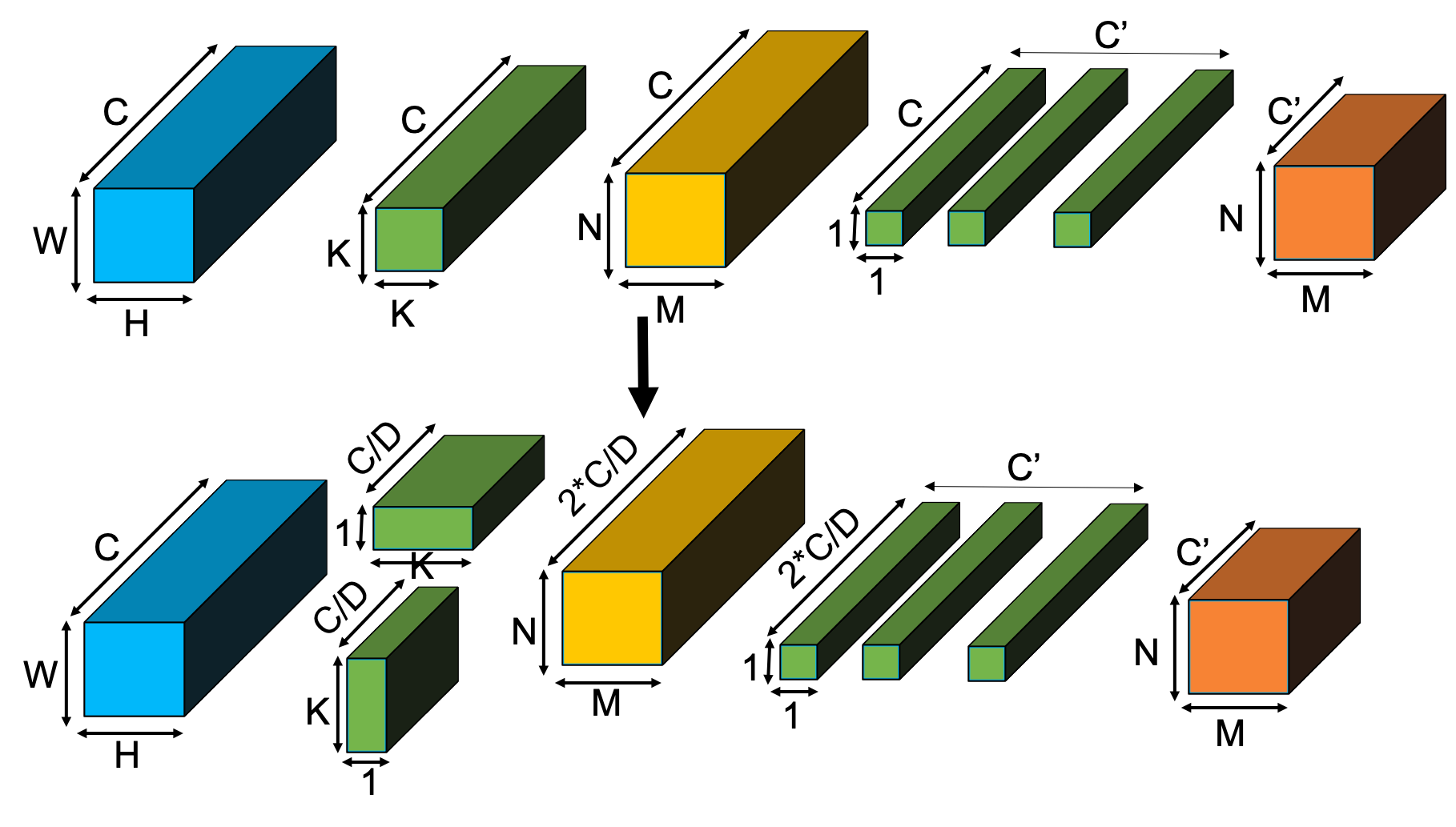}
        \caption{}
    \end{subfigure}%
    ~ 
    \begin{subfigure}[t]{0.5\textwidth}
        \centering
        \includegraphics[width=0.9\textwidth]{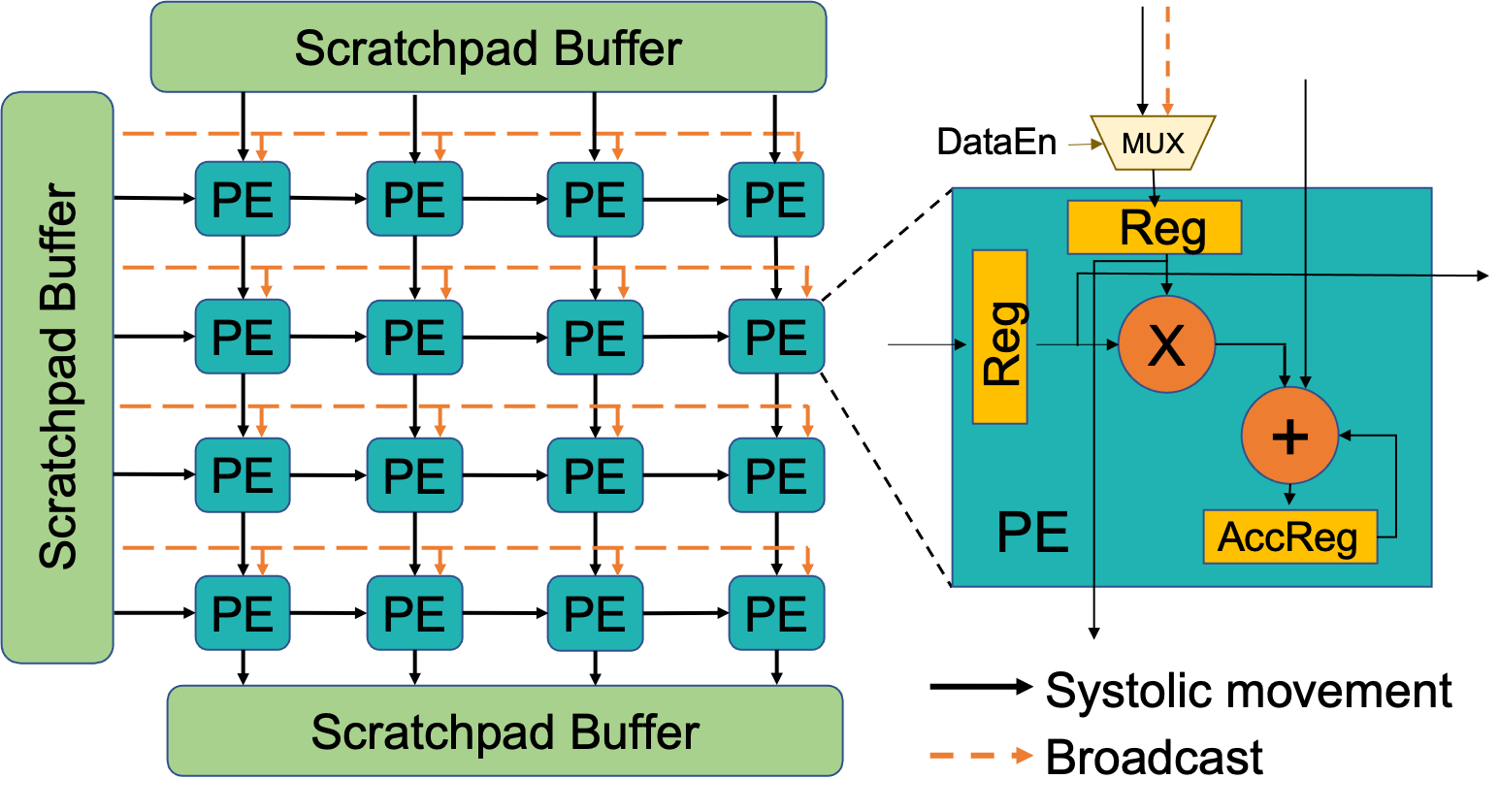}
        \caption{}
    \end{subfigure}
    \caption{(a) The proposed \fuseconv~operator (D=1 denotes \fusefull~and D=2 denotes \fusehalf) and (b) The proposed systolic-array architecture supporting \stos~dataflow}
    \label{fig:fuseconv_stos}
\end{figure}

\subsubsection{\fuseconv~is a systolic-algorithm}
\begin{figure}
    \centering
    \includegraphics[width=0.6\textwidth]{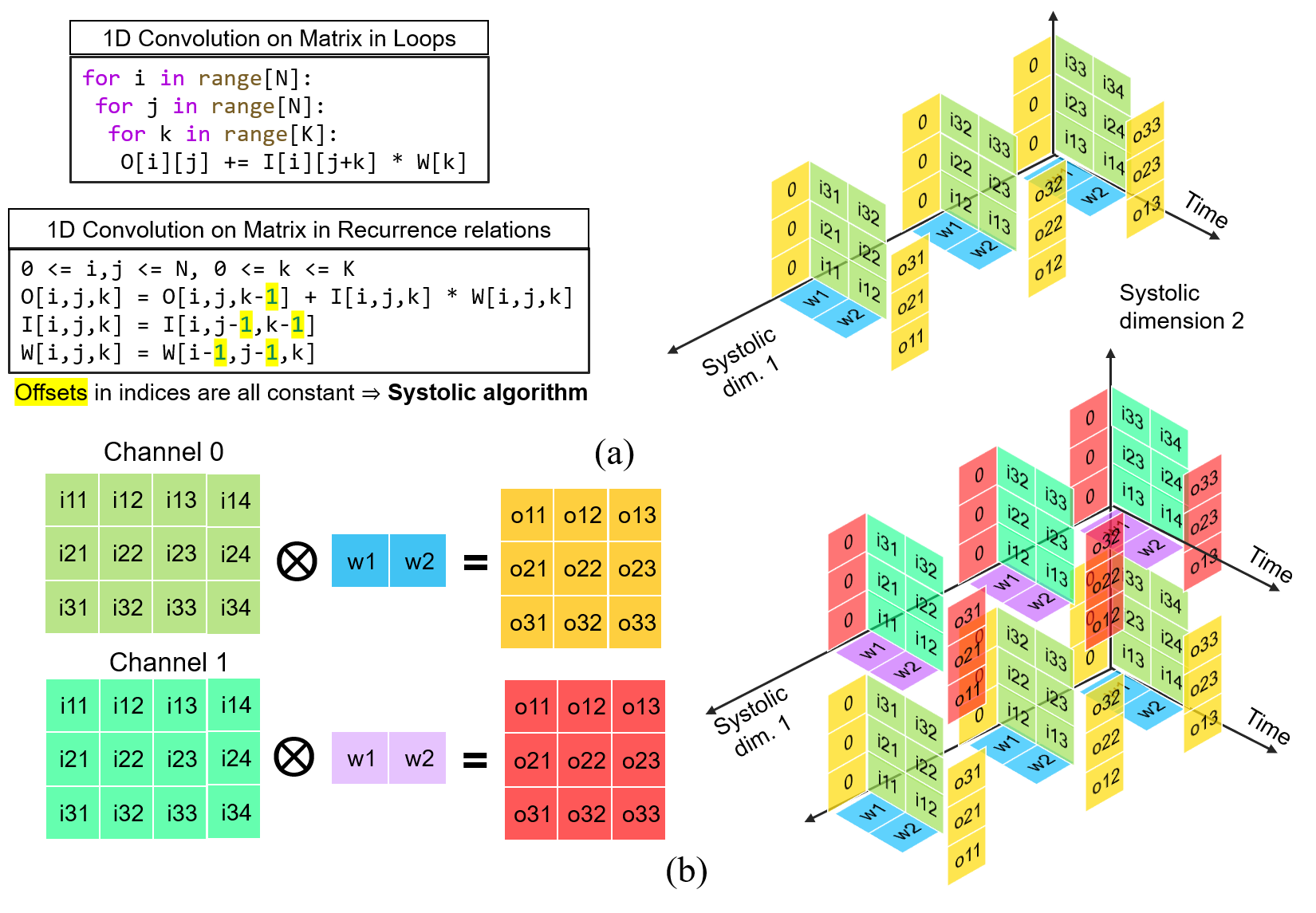}
    \caption{(a) Loop representation, recurrence relation and geometric representation of a 1D convolution; (b) Mapping multiple channels of 1D convolutions to the proposed systolic-array}
    \label{fig:1Dconv}
\end{figure}

\fuseconv~comprises of independent 1D convolutions which are clearly systolic algorithms \cite{quinton1984automatic}.
Indeed \cite{kung1982systolic} illustrates 7 different ways of mapping a 1D convolution on to a linear systolic array.
To confirm this intuition with the RIA formalism, in Fig.~\ref{fig:1Dconv} (a), we show the 1D convolution both as a loop and recurrence relations.
The recurrence relations satisfy all three requirements of an RIA. 
The other operation in a \fuseconv~layer, point-wise convolution, is a vector dot-product and is also a systolic algorithm.
Thus, \fuseconv~can be efficiently mapped to systolic arrays without requiring transformations such as im2col.

We recall again the sequence of claims so far.
2D convolution is not a systolic algorithm as it cannot be expressed in RIA form.
However, standard convolution with many channels and filters can be efficiently mapped on to a systolic array either by im2col transformations and filter reuse or by channel-wise computations.
When we factorize convolution channel-wise (depthwise) neither of these two tricks work.
Thus, depthwise convolution found in mobile bottleneck layers poorly utilizes systolic arrays.
However, when we further factorize convolution into the two spatial axes, the resultant 1D convolutions are again systolic algorithms and can be mapped efficiently on to an independent linear systolic row. 
It remains to be seen how such independent row-wise mapping can be implemented on two dimensional systolic arrays. 
It also remains to be seen if the parameter-efficient \fuseconv~achieves accuracy comparable to the baseline models with depthwise convolution. 
We discuss the former in the rest of this section, while the latter will be discussed in the experimental results. 


\subsection{Systolic-Array with \rtos~dataflow}

The choice of dataflow for a systolic architecture has a significant impact on its performance and energy-efficiency \cite{chen2016eyeriss, kwon2018co}. 
Since \fuseconv~is a systolic algorithm, avoids im2col transformations, and is entirely composed of 1D convolutions, it is at an advantage in being mapped to systolic arrays. 
Crucially, \fuseconv~requires that different 1D convolutions are mapped on to different rows of the systolic arrays. 
To enable this, we propose a novel dataflow, called \emph{Spatial-Tiled Output Stationary (\rtos)}.
In \rtos, 1D convolutions are computed on separate rows with the partial sums stationary in the processing elements. 
Hence the term \rtos, where the feature maps are \emph{tiled} along their spatial axes and scheduled with an output stationary dataflow. 
To support \rtos, we require an additional weight broadcast link to every row of the systolic array to feed the weight values to all PEs in that row. 
These broadcast links can co-exist with standard systolic dataflow which flows values activation values from the top to the bottom. 
We compute the area and power overhead of these additional links and find them to be nominal (see Section \ref{sec:evaluation}). 
Thus, the proposed systolic array architecture, shown in Figure \ref{fig:fuseconv_stos} (b), provides support for a runtime configurable dataflow, (i.e.) \rtos~with weight broadcast links for \fuseconv~and output- or weight-stationary dataflow for other operations. 
This design choice of allowing configurable dataflows has been well explored and shown to improve the efficiency of DNNs on systolic-arrays \cite{gholami2018squeezenext}.

In summary, we propose additional weight broadcast links on each row of the systolic array to map the 1D convolutions from \fuseconv.
This hardware-software co-design principle ensures that the systolic array is fully utilized while the network is parameter-efficient. 
In the following, we describe the process of mapping a given \fuseconv~network to a systolic array.

\subsection{Mapping \fuseconv~to systolic-arrays} 
\begin{figure}
    \center
    \includegraphics[width=\textwidth]{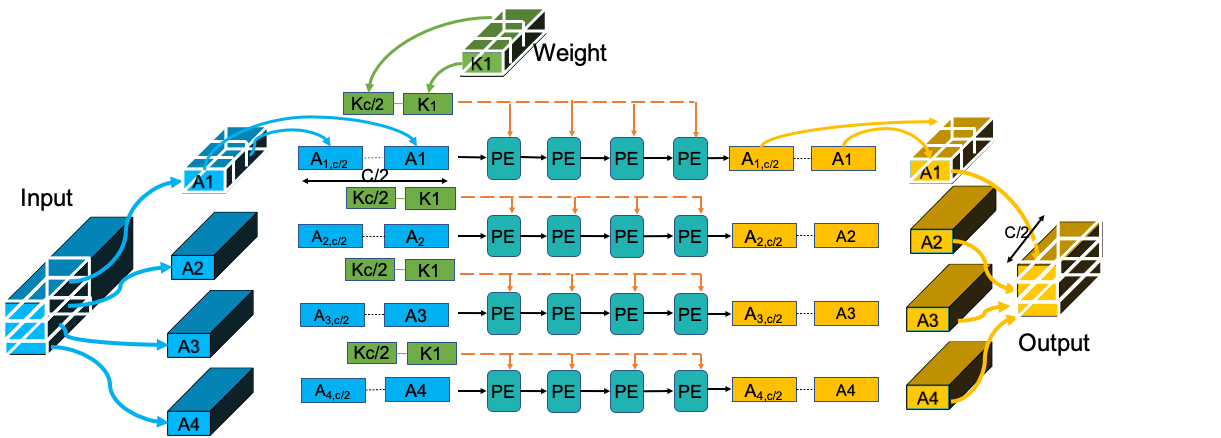}
    \caption{Mapping FuSe layers to the proposed systolic array architecture. \stos~dataflow with the additional weight broadcast enables mapping independent 1D convolutions to different rows enabling high-utilization of the hardware architecture.}
    \label{fig:mapping}
\end{figure}

We now concretely show the mapping of a \fuseconv~operation on to a systolic array with \stos~support.
Figure \ref{fig:mapping} illustrates the mapping for the \fusehalf~variant on to a systolic array of size $S\times S$. 
We show the mapping only for the row filters, which extends similarly to column filters. 
The input feature maps are first sliced across their spatial axes (rows, in this case) into $W$ rows, denoted $A_1$ through $A_W$.
The rows are further sliced across channels into $C/2$ channels denoted $A_{1,1}$ through $A_{1,C/2}$. 
Thus, there are $W\times C/2$ total 1D input slices that can be mapped to the rows of the systolic-array. 
Similarly, the independent channel weights are denoted $K_1$ to $K_{C/2}$ that are mapped to the row respective to their input feature map slice. 
The computation follows multiple folds wherein at each fold one weight-channel slice operates over one input channel slice to provide $S$ outputs. 
Finally, these outputs are concatenated across all the folds to form the output feature-map. 

Notably, there are two dominant design choices to map $W\times C/2$ slices to the rows of the systolic-array based on the available memory bandwidth: (a) \emph{Spatial-first mapping} and (b) \emph{Channels-first mapping}. 
In \emph{spatial-first mapping}, the slices from the same channel are given more precedence and are mapped to different rows of the systolic-array. 
Therefore, rows that operate on the same channel will operate with the same filter with lesser weight SRAM reads. 
However, this requires additional circuitry to broadcast the same weight filter to multiple rows. 
This mapping is suitable for bandwidth constrained systems.
In \emph{channels-first mapping}, the slices across channels are given more precedence and are mapped to the rows of the systolic array. Therefore, the rows work with distinct filters requiring multiple SRAM reads in the same clock cycle and do not require the additional broadcast circuitry. 
This mapping is suitable for systems that provide sufficient memory bandwidth. 
We can also map using a combination of both choices for small feature map inputs. 
For instance, in \emph{Channels-first mapping} if the channel size is smaller than the $S$ dimension, we can map the input feature maps across the remaining rows of the systolic array. 
We employ this hybrid mapping to capture the benefits of both design choices and to balance utilization for \fuseconv~operators with low number of channels. 

To illustrate the improvements in utilization for mapping \fuseconv~to the hardware with \rtos~dataflow, we visualize a 1D filter of 2 weights convolved with a 4x3 input in Figure \ref{fig:1Dconv} (b). 
Due to the broadcast link, at any time step the weight value is available along systolic dimension 1 (columns in the array).
Also different rows of the input are mapped along systolic dimension 2 (rows in the array), fully utilizing all processing elements.
We have explicitly shown the values of the input along the systolic dimension 1 due to \rtos.
In contrast to Figure \ref{fig:convsysalgo}(c) that shows depthwise convolutions only utilize a single column of the systolic array, the computation of \fuseconv~spans both systolic array dimensions, thereby achieving high utilization.
The span along dimension 1 increases as columns in the input increase, while the span along dimension 2 increases as rows in the input increase.\\


In summary, we proposed \fuseconv~and a systolic-array architecture supporting the novel \stos~dataflow as a HW/SW codesign solution to tackle the inefficiency of \dwsepconvs~on systolic-arrays. 
\fuseconv~networks offer an in-place replacement of \dwsepconvs~in existing mobile-efficient networks. 
However, \fuseconv~operator has fewer parameters than \dwsepconv~and thus we may expect an accuracy degradation. 
In the following section, we turn our attention to a methodology to train \fuseconv~networks by benefiting from existing \dwsepconv~networks.

\section{Neural Operator Scaffolding}
\label{sec:nos}
In this section, we study the third question we posed in the introduction on how to adapt the training procedure so that it better fits the paradigm of efficient \fuseconv~networks running with the \stos~dataflow. 
Towards this end, we propose Neural Operator Scaffolding (\nos) as a general technique to train networks with efficient operators benefiting from existing networks with more expensive operators.
We also combine NOS with search methods such as evolutionary algorithms and NAS to identify hybrid networks that combine \fuseconv~and depthwise separable convolution to trade-off model efficiency and accuracy. 

\subsection{Neural Operator Scaffolding (NOS)}
It is well understood that training deep models that are larger leads to more accurate models. 
Indeed, this scaling has a robust trend for Natural Language Processing (NLP) \cite{kaplan2020scaling}.
Similarly, it is also seen that while techniques like pruning are applicable in reducing model size, starting with a pruned network does not yield models of high accuracy. 
Thus, training models with a large number of parameters often appears to be a necessity to achieve high accuracy. 
For parameter-efficient operations like \fuseconv~this implies a challenge: As we will see in the experimental results, models where \fuseconv~is used as a drop-in replacement for depthwise separable convolution leads to higher than expected drops in accuracy (an average 2.1\% drop). 
The question then is can we benefit from the smaller model size and efficient inference supported by \fuseconv~but retain the high accuracy possible with more expensive operators.


Towards this end, we propose Neural Operator Scaffolding (\nos), as a training methodology. 
As suggested in the name, we ``scaffold'' the training of the efficient operators with more expensive operators by distilling information with adapter matrices. 
Adapter matrices have been explored in different settings such as parameter efficient transfer learning for downstream tasks and low-resource languages in NLP \cite{houlsby2019parameter,bapna2019simple, pfeiffer2020mad}, Neural Architecture Search \cite{cai2019once}, and distilling knowledge from a larger network to a smaller network \cite{jiao2019tinybert}. 
Similarly, distillation is a well studied technique to compress large teacher models into smaller student models. 
Distillation can happen at different levels - at the model level \cite{hinton2015distilling}, layer level \cite{sun2019patient,jiao2019tinybert}, or even functionality level \cite{touvron2020training}.
The key difference in NOS is the use of adapter matrices to distill knowledge between individual \textit{operators} - in this case a single convolutional kernel. 
More specifically, we distill knowledge from a depthwise convolution operator to its replacement \fuseconv~operator, such that the behavior of the depthwise convolution teacher operator is mimicked by the \fuseconv~student operator.
We see such distillation using adapter matrices as a general technique to search for efficient operators in building efficient and yet accurate models for the edge.



We now discuss the working of NOS with some notation. 
Consider two operators $T$ and $S$ (for teacher and student), operating on an input tensor $I$, producing $O_T$ and $O_S$ tensors, i.e., $O_T = T(I)$ and $O_S = S(I)$.
For instance $T$ can denote depthwise convolution while $S$ can be \fuseconv.
The key idea of NOS is to relate $T$ and $S$ operators with an adapter function $A$, i.e., $S = A(T)$.
In our specific case, we choose $A$ such that the weights that characterise $T$ and $S$ are related with linear projections. 

\begin{figure}
    \centering
    \includegraphics[width=\textwidth]{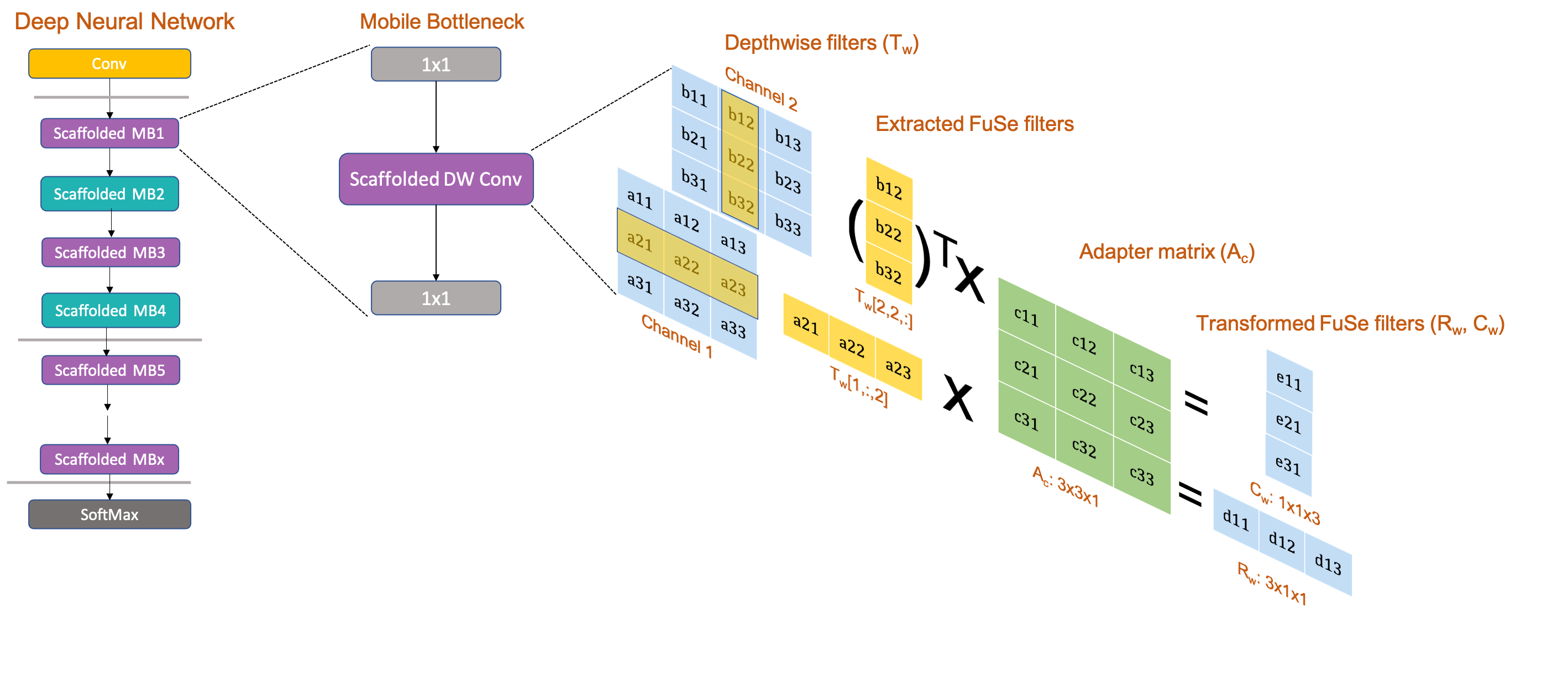}
    \caption{Illustration of Neural Operator Scaffolding. Adapter matrices are added to every mobile-bottleneck layer that learns to distill knowledge from ``expensive'' depthwise operators to ``cheap'' \fuseconv~operators.}
    \label{fig:nos}
\end{figure}

In particular, consider $T$ as a depthwise convolution with a kernel $T_w$ of size $2 \times 3 \times 3$ as shown in Figure \ref{fig:nos}, with the dimensions denoting channel, rows, and columns, respectively.
Now consider a drop-in replacement of this with a \fuseconv~operator denoted $S$, which has one row filter $R_w$ of size $3 \times 1 \times 1$ for the first channel and one column filter $C_w$ of size $1 \times 1 \times 3$ for the second channel. 
Our linear adapter in this case, maps the weights of two kernels. 
In particular, $R_w = A_r \times T_w[1, :, 2]$, and similarly $C_w = A_c \times T_w[2, 2, :]$, where $A_r$ and $A_c$ are matrices of size $3 \times 3$.
In this work, we use the same matrix for both $A_r$ and $A_c$, i.e., row and column filters have the same adapter matrices. 
Further, we share these adapter matrices across \textit{all} filters in the same layer. 
Thus, there are only $K^2$ additional trainable parameters in scaffolding a depthwise convolution layer to \fuseconv.
In the example from Figure~\ref{fig:nos}, $K$ is 3, and thus we have 9 additional trainable parameters in $A_r$ along with the 18 trainable parameters in the $T_w$ for depthwise separable convolution.
Such adapter matrices can be added to each bottleneck layer in a network, to obtain the scaffolded network. 

How do we train this scaffolded network?
First we require that a \textit{teacher} network with the more expensive \dwsepconvs~is already trained on the task. 
In addition, a student scaffolded network with shared adapter matrices in each layer is created.
One option is to train the scaffolded network as is, i.e., training both the depthwise convolution weights and adapter matrices. 
However, similar to once-for-all (OFA) \cite{cai2019once}, we observe empirically that it is better to train networks with combinations of depthwise and \fuseconv~layers. 
That is, at each training step, all the scaffolded layers of the student network are randomly chosen to be either \dwsepconv~or \fuseconv. 
The input batch is then propagated through this sampled network.
For layers with only \dwsepconv, only the filter weights are updated through backpropagation, and for layers with \fuseconv, both the filter weights and the adapter matrices are updated.
In addition to the standard loss, we also have a knowledge distillation loss  based on the output of the teacher network (all \dwsepconv) on the same batch of inputs.
Specifically, the loss is the difference in the logits or soft labels produced by the student and teacher networks as proposed in the original work on knowledge distillation \cite{hinton2015distilling}. 

After training, we would have the filter kernels of depthwise convolution and the adapter matrices, which can be collapsed to obtain the filter weights of the corresponding \fuseconv. 
In other words, we remove the scaffold of the adapter matrices and run inference only on the efficient \fuseconv~layers. 
Thus, \nos~is only a training procedure - we could either train \fuseconv~layers directly or with scaffolding from the more expensive depthwise convolution. 
We compare these approaches in the experimental section to establish the value of scaffolded training. 

 


\subsection{Combining \nos~with search exploration strategies}
\nos~as proposed above generates a network with only \fuseconv.
However, \nos~can be combined with a search procedure that searches for hybrid networks that combine \dwsepconvs~and \fuseconv.
For a network with $N$ mobile-bottleneck layers, we have $2^N$ possible choices of hybrid networks.
These choices would enable an accuracy-latency trade-off space which can be explored depending on the hardware constraints and application requirements.


One approach to search for hybrid networks is optimization with Evolutionary Algorithms (EAs) \cite{real2017large}. 
These algorithms start with a population of networks and apply genetic operations (mutation, crossover) to iteratively update the population of networks. 
The fittest networks are retained where fitness could be based on accuracy and latency metrics. 
Accuracy and latency measurements can be slow when considering many networks, and thus approximate cost models are often used \cite{cai2018proxylessnas,cai2019once, ganesan2020case}.


Beyond searching for hybrid networks, we can also search for architectures by changing other features such as number of bottleneck layers and the kernel sizes. 
Clearly these choices also affect the accuracy-latency trade-off and such search is referred to as Neural Architecture Search (NAS).
We explore combining \nos~with one such NAS algorithm namely Once-For-All (OFA) \cite{cai2019once}.
This requires us to add elastic training schedules for each dimension - width, depth, and kernel sizes, as done in OFA \cite{cai2019once}.\\


In summary, we propose \nos~as a training methodology to train more accurate \fuseconv~networks by distilling knowledge from more expensive operators with adapter matrices. 
This scaffolding can be combined with EA or NAS to trade-off accuracy and latency by searching a large set of hybrid networks with other architectural choices.
We evaluate \nos~individually and in combination with EA and NAS in the experimental section.

\section{Experimental Methodology} 
\label{sec:experimental}
\noindent
In this section, we share the methodology used for experimentally evaluating \fuseconv~networks. 
We begin first with a presentation of SCALE-Sim-FuSe which extends the SCALE-Sim simulator \cite{samajdar2018scale} for systolic arrays. 
We then discuss the training setup. 

\subsection{SCALE-Sim-FuSe}
To evaluate the latency of executing \fuseconv~networks on systolic arrays, we use SCALE-Sim \cite{samajdar2018scale} by adding support for \stos~dataflow. 
SCALE-Sim is a cycle-accurate behavioral simulator to model systolic arrays enabling quick design space exploration of hardware design points.
To enable this exploration, the simulator is highly configurable allowing users to choose across dataflow, number of Processing Elements (PEs), the size of buffer sizes, etc. 
In addition, it allows users to (a) log cycle-by-cycle metrics, (b) model bandwidth demands due to memory constraints, (c) generate SRAM and DRAM traffic traces and compute Processing Element (PE) utilization and execution times. 

However, SCALE-Sim models only 2D spatial convolutions and linear layers, and does not model \dwsepconvs~or \fuseconv. 
Thus, we enhance SCALE-Sim to support both \dwsepconv~and \fuseconv~networks. 
This support involves allowing channel-wise filters to operate independently for \dwsepconvs~and supporting the \stos~dataflow for \fuseconv. 
The enhanced version is open-sourced at \url{https://github.com/iitm-sysdl/SCALE-Sim-FuSe}.
Table \ref{table:sysconfig} lists the configurations of the systolic-array used for evaluating the performance of \fuseconv. These configurations are used throughout unless otherwise explicitly mentioned. 

\begin{table}[!h]
\centering
\vspace{0pt}
\scalebox{0.80}{
\begin{tabular}{l|cc}
\toprule
Operating frequency & 1GHz \\
\midrule
Array dimensions & $16\times 16$ \\
\midrule 
Dataflow & Output-Stationary and \rtos \\
\midrule
Input feature map SRAM size & 64KB \\
Weight SRAM size & 64KB \\
Output feature map SRAM size & 64KB \\
\bottomrule
\end{tabular}
}
\caption{System Configuration}
\label{table:sysconfig}
\end{table}

\subsection{Design Overheads of \stos~dataflow}
Before we present results on the analysis of \fuseconv~networks, we evaluate the additional hardware cost required to support \fuseconv~- namely the design overhead of supporting the \stos~dataflow.
To this end, we implemented systolic arrays of different diemnsions on Bluespec System Verilog \cite{nikhil2004bluespec} and synthesized them on a proprietary library in 22nm using Synopsys Design compiler. 
Two variants with and without \rtos~were implemented and their areas and powers, measured through Synopsys Design Compiler, were compared. 
We report the percentage overheads in both area and power across 4 different sizes in Table~\ref{table:overheads}.
The overheads increase the array sizes, but remain acceptably small at 5.2\% area overhead and 9.2\% power overhead for a $64 \times 64$ systolic array.
For the specific case of inference on the edge, we do not expect the synthesis of arrays larger than this size. 
Thus, we conclude that the design overheads of \stos~acceptably small.

\begin{table}[!h]
\centering
\vspace{0pt}
\scalebox{0.80}{
\begin{tabular}{l|c|c}
\toprule
Array Dimension & \begin{tabular}[c]{@{}l@{}}Area\\overhead (in \%)\end{tabular}  & \begin{tabular}[c]{@{}l@{}}Power\\overhead (in \%)\end{tabular} \\ 
\toprule
8x8                                                  & 3 & 6.2 \\
16x16                                                & 3.2 & 6.7 \\
32x32                                                & 4.5 & 6.4 \\
64x64                                                & 5.2 & 9.2 \\
\toprule
\end{tabular}
}
\caption{Area and Power Overheads of varying Systolic-Array dimensions with \rtos~dataflow support}
\label{table:overheads}
\end{table}

\subsection{Training methodology} 
We now describe the training methodology for training \fuseconv~networks - individually and with \nos.
We refer to the former as in-place replacement (of \dwsepconv).
We report results on MobileNet V1, V2, V3-Small, V3-Large, MNasNet-B1 \cite{howard2017mobilenets,sandler2018mobilenetv2,howard2019searching,tan2019mnasnet}. 
These models are particularly designed for edge devices such as mobile phones, and are amongst the most efficient models used today.
While the MobileNet family is obtained by a combination of operator design (eg. bottleneck layers) and NAS, the MNasNet model is based purely on NAS. 
In all our results, we train and evaluate models on the ImageNet dataset \cite{deng2009imagenet}.

\subsubsection{In-place replacement}
In this method, we replace all the \dwsepconv~layers  of MobileNet (V1, 2, 3) \cite{howard2017mobilenets,sandler2018mobilenetv2,howard2019searching} and MnasNet-B1 \cite{tan2019mnasnet} with \fuseconv~layers, train them, and evaluate the networks for accuracy on ImageNet dataset \cite{deng2009imagenet}. 
We use PyTorch \cite{paszke2019pytorch} to describe and train the models. 
We use RMSProp as the optimizer with an initial learning rate of $0.016$ and momentum of $0.9$. 
The learning rate schedule follows an exponential decay of $0.97$ every $2.4$ epochs. 
We maintain an Exponential Moving Average (EMA) of the weights with a decay factor of $0.999$. 
The weight-decay is $1e-5$. 
We use FP16 for training and train the models for 350 epochs on 8V100s or 4P100s with a batch size of 128 per GPU. 



\subsubsection{Neural Operator Scaffolding}
In \nos, we distill knowledge from pretrained networks which have \dwsepconv, specifically from MnasNet-B1 and MobileNetV3-Large - two efficient networks which achieve the highest accuracy in ImageNet dataset. 
We use a hyperparameter schedule adapted from once-for-all \cite{cai2019once}. 
We use SGD as the optimizer with an initial learning rate of $0.03$ and momentum $0.9$. 
We follow a cosine schedule in the learning rate.
The weight decay was $3e-5$. 
We use Dropout with probability $0.1$ and label smoothing with a factor of $0.1$. 
We set batch-normalization momentum and epsilon at $0.1$ and $1e-5$, respectively. 
We train the model for 350 epochs with the scaffolding. 
All models were trained on 8 V100 GPUs with a batch size of 128 per GPU using FP16 precision. 

As discussed, \nos~can be combined with evolutionary algorithms (EAs) and Neural Architecture Search (NAS). 
For EA, we adapt the algorithm presented in \cite{real2017large}.
We choose a population size of $100$, mutation probability of $0.1$, and the ratio of parent and mutation of $0.25$. 
The search algorithm has a bound of $100$ iterations. 
For NAS, we use once-for-all \cite{cai2019once} and replicate the progressive shrinking algorithm along three elastic dimensions - kernel sizes ($3 \times 3$, $5 \times 5$, or $7 \times 7$), depth (2, 3, or 4), and width (3, 4, or 6) and scaffold adapter matrices across kernel sizes to enable \fuseconv.
The networks are trained for 1000 epochs on 8 V100 GPUs using FP16 precision.

\section{Results} 
\label{sec:evaluation}

In this section, we share the findings of experimentally evaluating \fuseconv~networks on the systolic array hardware supporting \stos~dataflow. 
We first compare \fuseconv~as a drop-in replacement for \dwsepconv~in terms of accuracy and latency for different efficient computer vision networks.
We also analyze latency numbers in detail - layer-wise contribution, effect of scaling systolic arrays, and impact of bandwidth (both DRAM and SRAM). 
We then evaluate the performance of NOS - individually, and with evolutionary search and Neural Architecture Search. 

\subsection{Speedup of \fuseconv}
We now evaluate the latency benefits of \fuseconv~measured with SCALE-Sim-FuSe. 
Recall that we argued that FuSeConv with \stos~can fully utilize systolic arrays. 
In line with this theoretical observation, we report significant speedups in this section.
We also analyze the latency values to understand components which contribute to the latency the most.
All evaluations are done on the mobile-efficient networks described in Table \ref{table:networks}. 
The more efficient \fusehalf~operator is used throughout this evaluation unless explicitly specified otherwise.

Figure \ref{fig:speedup_layerwise}(a) reports the latencies of baseline mobile-efficient networks (without \fuseconv) and \fuseconv~networks executed with Output Stationary (OS) and Weight Stationary (WS) \cite{chen2016eyeriss} dataflows and \rtos~dataflow, respectively, on $16 \times 16$ systolic arrays.
We report significant speedups of $7.01-9.36\times$ on the \fusehalf~variants and $4.15-5.05\times$ on the \fusefull~variants with \rtos~relative to the Output Stationary baseline. 
This reinforces our finding that \dwsepconvs~are inefficient to execute on systolic arrays and factorizing them to 1D convolution leads to significantly improved performance. 
It is noteworthy that the \fusefull~variant is able to achieve significant speedups despite higher MACs and parameters. 

\subsubsection{Layerwise analysis}
\begin{figure}
    \centering
    \includegraphics[width=1.1\textwidth]{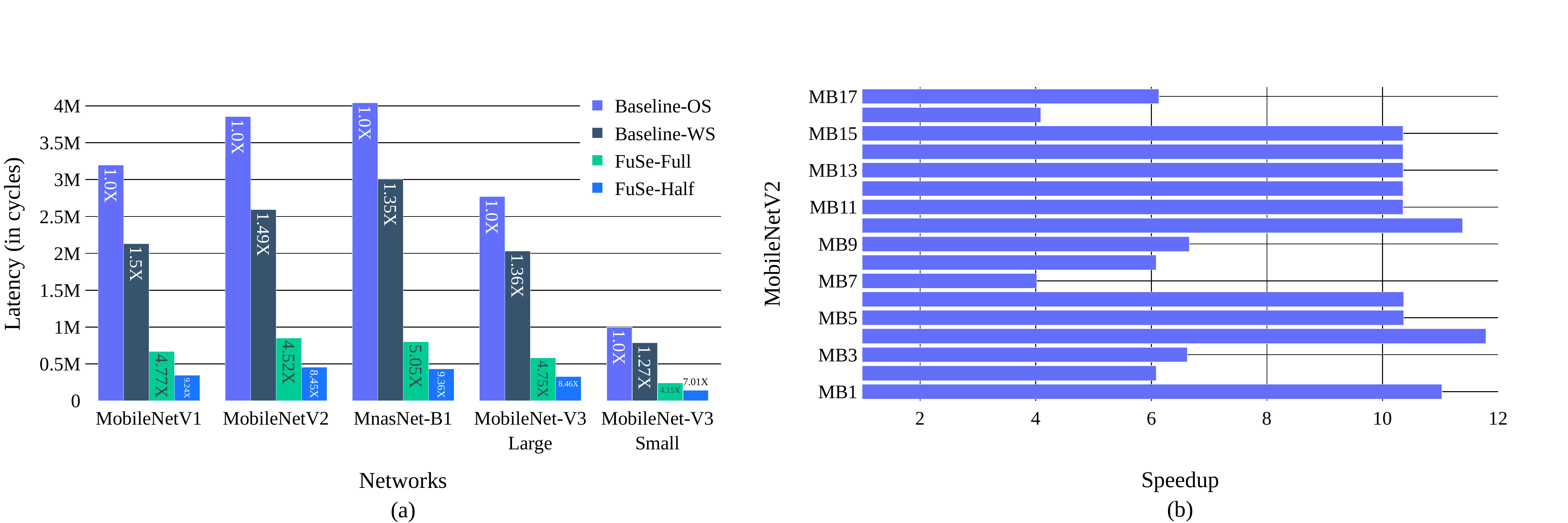}
    \caption{Experimental results evaluating the latency benefits of FuSeConv: (a) Latency estimates on $16\times 16$ systolic arrays, and (b) Layerwise speedups for MobileNetV2. \fuseconv~with \stos~dataflow significantly improves the performance of efficient networks on systolic arrays.}
    \label{fig:speedup_layerwise}
\end{figure}

Figure \ref{fig:speedup_layerwise}(b) reports the layerwise speedup for the \fusehalf~variant of MobileNetV2. 
The speedup ranges from $4$ to $11\times$, with some middle and later layers exhibiting lower relative speedups. 
This is primarily due to the size of the layers - the final and intermediate smaller layers are too small to fully utilize the systolic array. 
We discuss this in more detail when analyzing the hardware utilization with each layer.

\subsubsection{Operator-wise analysis}
\begin{figure}
    \centering
    \includegraphics[width=1.1\textwidth]{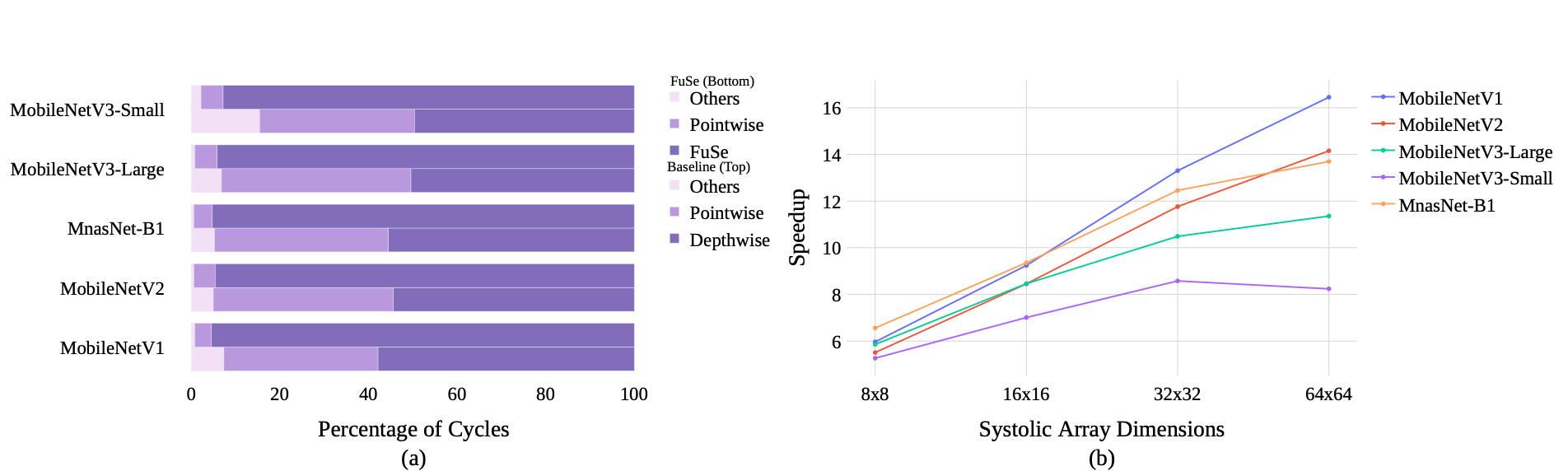}
    \caption{Operator-wise and scaling analysis \fuseconv: (a) Latency distribution of operators for baseline and \fuseconv~ and (b) Ablation study. \fuseconv~optimizes the common-case depthwise operator and exhibits strong performance scaling for increasing systolic array sizes.}
    \label{fig:scaling_percentage}
\end{figure}

We report the latency distribution across operators for all the networks. 
As shown in Figure \ref{fig:scaling_percentage}(a), we see that the latency of baseline networks is dominated by depthwise convolutions (>90\%) making it the common case for optimization on the edge.
With the \fuseconv, the latency distribution shifts towards pointwise and other convolutions and becomes more balanced where the \fuseconv~operation accounts for a smaller fraction of the overall latency (on average <50\%). 
This establishes that the combination of \fuseconv~and the \rtos~dataflow optimizes the common case.
Further improvements in performance can come from speeding up pointwise convolutions for instance with grouped convolutions \cite{krizhevsky2012imagenet}.

\subsubsection{Systolic array utilization}
\begin{figure}
    \centering
    \includegraphics[width=1.1\textwidth]{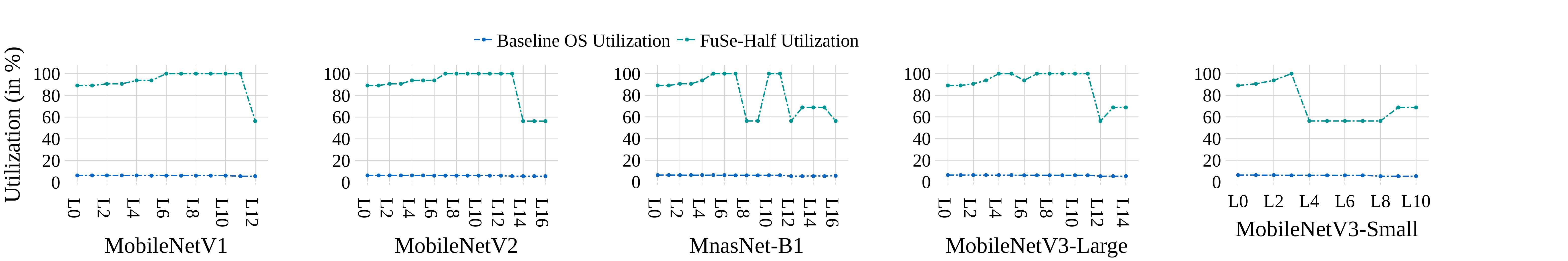}
    \caption{Contrasting the utilization of \fuseconv~layers with \dwsepconv~layers of efficient-networks on a 16x16 systolic-array. \fuseconv~exhibits high-utilization due to the massive parallelism enabled by \stos~dataflow.}
    \label{fig:utilization}
\end{figure}

To understand the speedup observed with \fuseconv, we report the systolic array utilization of all the mobile bottleneck layers (depthwise or \fusehalf~convolution sandwiched between two $1\times 1$ pointwise convolution) for all the networks in Figure \ref{fig:utilization}. 
We observe that the utilization of layers with \fusehalf~operators are very high (56-100\%), while the corresponding baseline networks exhibit a very poor utilization (5-6\%). 
This again reinforces our finding that \dwsepconv~benefits neither from data reuse or channel-wise computation in utilizing all columns of the systolic array.
Within \fuseconv~layers, we observe trends that corroborate our findings in the layerwise breakdown. 
First, the final bottleneck layers exhibit relatively lower utilization (~50-60\%) due to extremely small input feature map sizes (as low as $7\times 7$). 
Second, some of the middle layers also exhibit lower utilization due to input-feature map padding creating halo regions. 
Finally, MobileNetV3-Small is a very tiny network with ultra-low parameters and MACs, and it's not able to sufficiently saturate a $16\times 16$ array despite the tremendous parallelism offered by \rtos~dataflow.

\subsubsection{Bandwidth utilization}
\begin{figure}
    \centering
    \includegraphics[width=1\textwidth]{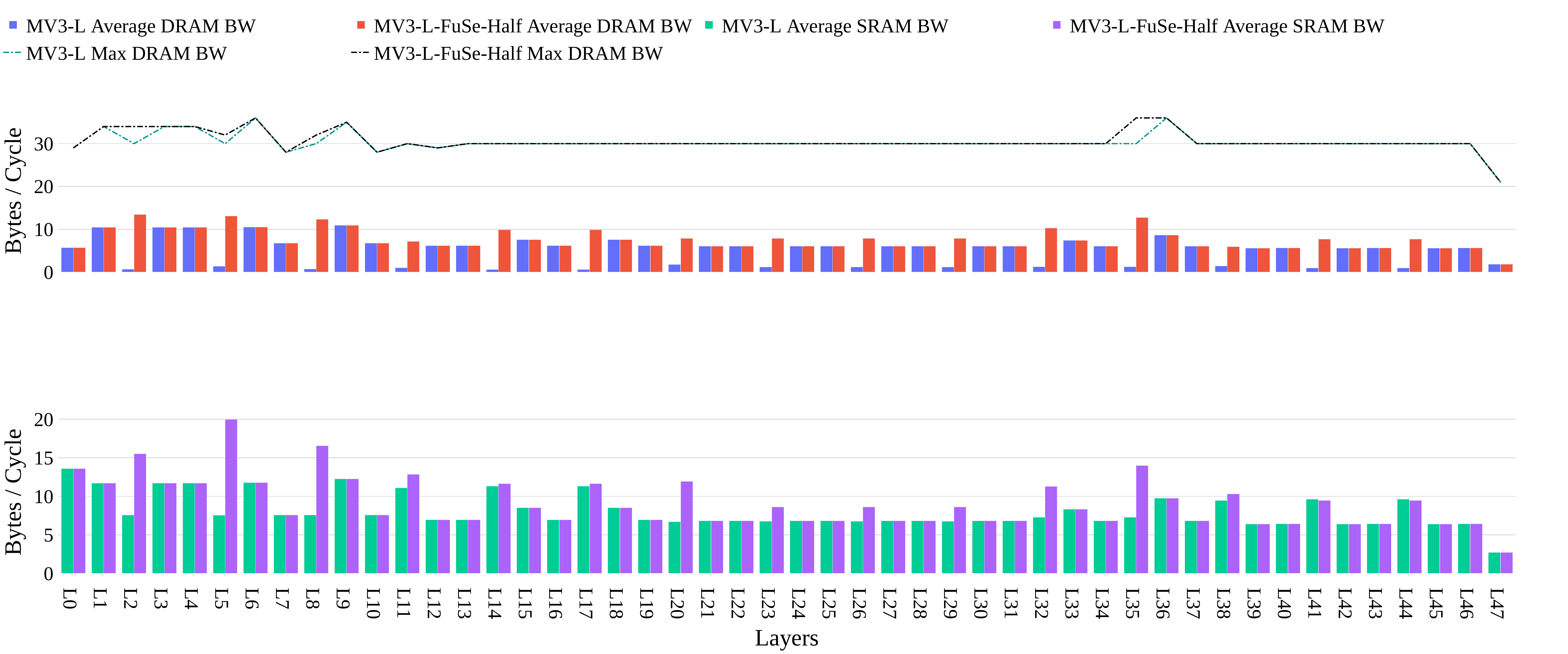}
    \caption{Layerwise DRAM and SRAM bandwidths for baseline and \fuseconv~variants of MobileNetv3-Large. The bandwidth consumed by \fuseconv~layers are not substantially higher than other operators (such as pointwise) establishing that the existing memory subsystem can support \fuseconv~ comfortably.}
    \label{fig:bandwidth_percentage}
\end{figure}

In order to understand the bandwidth requirements of \fuseconv~and the \rtos~dataflow, we report the SRAM and DRAM bandwidths (average and maximum) utilized by each layer of MobileNetV3-Large (with and without \fusehalf) in Figure \ref{fig:bandwidth_percentage}. 
We observe that \fusehalf~layers (every $3^{rd}$ layer) utilizes a higher operating average bandwidth than the depthwise convolution layers. 
This is because the \rtos~dataflow enables massive parallelism thus requiring more data requests to both the SRAM and the DRAM. 
However, when compared to the non-FuSe layers (pointwise and spatial convolutions), this increase in average bandwidth for \fusehalf~layers is not substantial. 
Particularly, the maximum bandwidth requirements for DRAMs are similar for both depthwise convolution layers.
Thus, to support \fuseconv~networks, the peak DRAM bandwidth required is similar to other standard operators.


\subsubsection{Scaling with systolic array size}
We report results (refer Figure \ref{fig:scaling_percentage}(b)) of the effect of the systolic array size on the speedup of \fuseconv~networks.
We observe that the speedup continues to increase for higher systolic array sizes. 
Interestingly, smaller networks such as MobileNetV3-small exhibit an overall weaker scaling where the speedup peaks at a systolic array size of $32 \times 32$. 
Such analysis may be useful in designing the systolic array size for accelerators on the edge. 

In summary, we have demonstrated that the \fuseconv~operator is significantly more efficient than \dwsepconv~on systolic arrays.
This leads to about $8\times$ speedup on average for popularly deployed mobile efficient networks.

\subsection{Accuracy of \fuseconv~with in-place replacement} 
We begin with analyzing the accuracy of in-place replacement of \dwsepconvs~with \fuseconv~for different efficient models - MobileNet v1, v2, v3-Small, v3-Large, and MNasNet-B1.
We report all our results on the ImageNet dataset \cite{deng2009imagenet} with the standard train and test splits.

For each of these efficient networks, we study different variants for in-placement replacement. 
The Full and Half variants are as described in Section~\ref{sec:fuseconv_stos}.
The Full variant has about twice the number of MACs as the Half variant. 
For both these variants, we also study two other variants denoted with 50\%, where only half the bottleneck layers are replaced with \fuseconv. 
Here the layers to be replaced are chosen greedily based on the impact on latency.
These 50\% variants will have fewer parameters and MACs than the respective parent variants. 
For instance, the Full-50\% variant has about one-third fewer MACs than the Full variant.

For each of these variants, we train models on ImageNet and report the accuracy values in Table \ref{table:networks}. 
We make the following observations on the results: 
\begin{itemize}
    \item For the older models, MobileNet-V1 and V2, some of the \fuseconv~variants are more accurate than baseline models.
    For instance, for MobileNet-v1, replacing the depthwise operators with FuSe-Half leads to a more accurate model with fewer MACs and parameters.
    \item On average across all models, the drop in accuracy for the FuSe-Full model is 0.3\%.
    But the use of the FuSe-Full model is not feasible for models on the edge given that both MACs and parameters are larger than the baseline models. 
    \item On average across all models, the drop in accuracy for the FuSe-Half model is 2.3\%. 
    This a reasonably large drop and suggests the use of other methods to make up for this loss. 
    \item When using 50\% variants, the average drop in accuracy for FuSe-Full and FuSe-Half variants reduces to <0.2\% and 0.8\%, respectively. 
\end{itemize}


     
\begin{table}[!h]
\centering
\vspace{0pt}
\scalebox{0.80}{
\begin{tabular}{lccc}
\toprule
Network & \begin{tabular}[c]{@{}l@{}}ImageNet\\accuracy\end{tabular}  & \begin{tabular}[c]{@{}l@{}}MACs\\(millions)\end{tabular} & \begin{tabular}[c]{@{}l@{}}Params\\(millions) \end{tabular} \\ 
\toprule
MobileNet-V1\cite{howard2017mobilenets}              &  70.60 & 589  & 4.23 \\
MobileNet-V1 FuSe-Full                               &  72.86 & 1122 & 7.36 \\
MobileNet-V1 FuSe-Half                               &  72.00 & 573  & 4.20 \\
MobileNet-V1 FuSe-Full-50\%                          &  72.42 &  764 & 4.35 \\
MobileNet-V1 FuSe-Half-50\%                          &  71.77 &  578 & 4.22 \\
\midrule
MobileNet-V2\cite{sandler2018mobilenetv2}            & 72.00  & 315 & 3.50  \\
MobileNet-V2 FuSe-Full                               & 72.49  & 430 & 4.46 \\
MobileNet-V2 FuSe-Half                               & 70.80  & 300 & 3.46 \\
MobileNet-V2 FuSe-Full-50\%                          & 72.11  & 361 & 3.61 \\
MobileNet-V2 FuSe-Half-50\%                          & 71.98  & 305 & 3.49 \\
\midrule
MnasNet-B1\cite{tan2019mnasnet}                      &  73.50  & 325 & 4.38 \\
MnasNet-B1 FuSe-Full                                 &  73.16  & 440 & 5.66 \\
MnasNet-B1 FuSe-Half                                 &  71.48  & 305 & 4.25 \\
MnasNet-B1 FuSe-Full-50\%                            &  73.52  & 361 & 4.47 \\
MnasNet-B1 FuSe-Half-50\%                            &  72.61  & 312 & 4.35 \\
\midrule
MobileNet-V3 Small \cite{howard2019searching}        &  67.40  & 66 & 2.93 \\
MobileNet-V3 Small FuSe-Full                         &  67.17  & 84 & 4.44 \\
MobileNet-V3 Small FuSe-Half                         &  64.55  & 61 & 2.89 \\
MobileNet-V3 Small FuSe-Full-50\%                    &  67.91  & 73 & 3.18 \\
MobileNet-V3 Small FuSe-Half-50\%                    &  66.90  & 63 & 2.92 \\
\midrule
MobileNet-V3 Large \cite{howard2019searching}        & 75.20   & 238 & 5.47 \\
MobileNet-V3 Large FuSe-Full                         & 74.40   & 322 &10.57 \\
MobileNet-V3 Large FuSe-Half                         & 73.02   & 225 & 5.40 \\
MobileNet-V3 Large FuSe-Full-50\%                    & 74.50   & 264 & 5.57 \\
MobileNet-V3 Large FuSe-Half-50\%                    & 73.80   & 230 & 5.46 \\
\hline
\toprule
\end{tabular}
}
\caption{ImageNet performance, parameters, and MACs of DNNs used to evaluate FuSeConv.}
\label{table:networks}
\end{table}

In summary, our results show that the in-place replacement with FuSe-Half operator leads to a drop in accuracy large enough to require further optimization. 
Replacing only half the layers improves the accuracy, but further optimizations are required for training networks with the efficient operator.

\subsection{Accuracy of \fuseconv~with \nos}
We proposed \nos~to improve the training of efficient operators by distilling knowledge from more expensive operators.
We apply this procedure for the two most accurate networks, namely MobileNetV3-Large and MnasNet-B1. 
We observe that the accuracy of these two networks increases by 0.8\% and 1.5\% respectively, when training with \nos. 
These improvements account for 37\% and 74\% of the gap between the \fuseconv~and \dwsepconv~variant. 
This demonstrates the effectiveness of operator scaffolding - without any change in the network architecture or the inference mechanism we obtain improved accuracy results. 
When comparing the \fuseconv~variant of MobileNetV3-Large trained with \nos~with MobileNetV3-Small we observe that, the \fuseconv~variant is 5.8\% more accurate while being 2.3$\times$ faster.
Therefore, instead of downgrading from MobileNetV3-Large to MobileNetV3-Small for constrained edge scenarios, we can use the \fusehalf~variant of MobileNetV3-Large as a strong state-of-the-art model.

\begin{figure}
    \centering
    \includegraphics[width=0.8\textwidth]{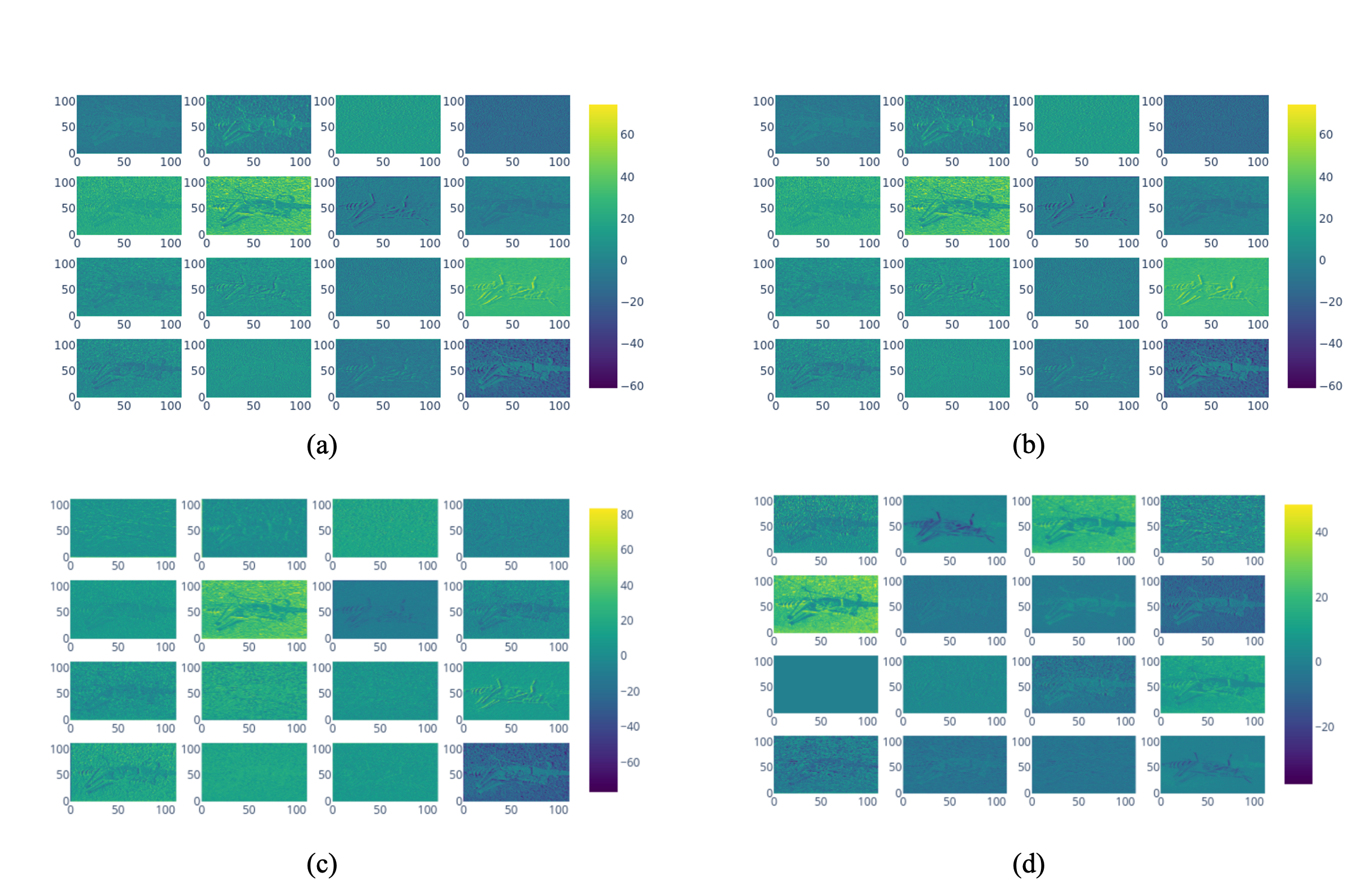}
    \caption{Feature map visualization of $3^{rd}$ mobile bottleneck layer: (a) and (b) MobileNetV3-Large and its \fusehalf~variant trained with NOS, and (c) and (d) The same networks trained without NOS. The feature map visualizations are identical for \nos~while different for standard training, illustrating the effectiveness of knowledge distillation with \nos.}
    \label{fig:fmap_vis}
\end{figure}

We hypothesize that training with \nos~is effective because the \fuseconv~operator is able to distill the knowledge from the respective \dwsepconv~operator. 
To test this hypothesis, we visualize the output of the third mobile bottleneck layer for a specific input image of MobileNetV3-Large trained with and without \nos. 
We show this in Figure \ref{fig:fmap_vis} in four parts. 
(a) and (b) are the outputs for baseline \dwsepconv~and \fusehalf~variants with \nos~training, while (c) and (d) are the outputs for the same networks with in-place replacement (without \nos). 
From the figures, it is clear that (a) and (b) have high similarity indicating effective knowledge distillation from the \dwsepconv~operator to the \fuseconv~operator. 
On the other hand, (c) and (d) are very different suggesting that the \fuseconv~operator is quite distinct from the \dwsepconv~operator it is replacing.
This ``scaffolding'' is responsible for the higher accuracy we observe with \nos.


\subsection{Accuracy of \fuseconv~with \nos~and Evolutionary Search}

\begin{figure}
    \centering
    \includegraphics[width=1.1\textwidth]{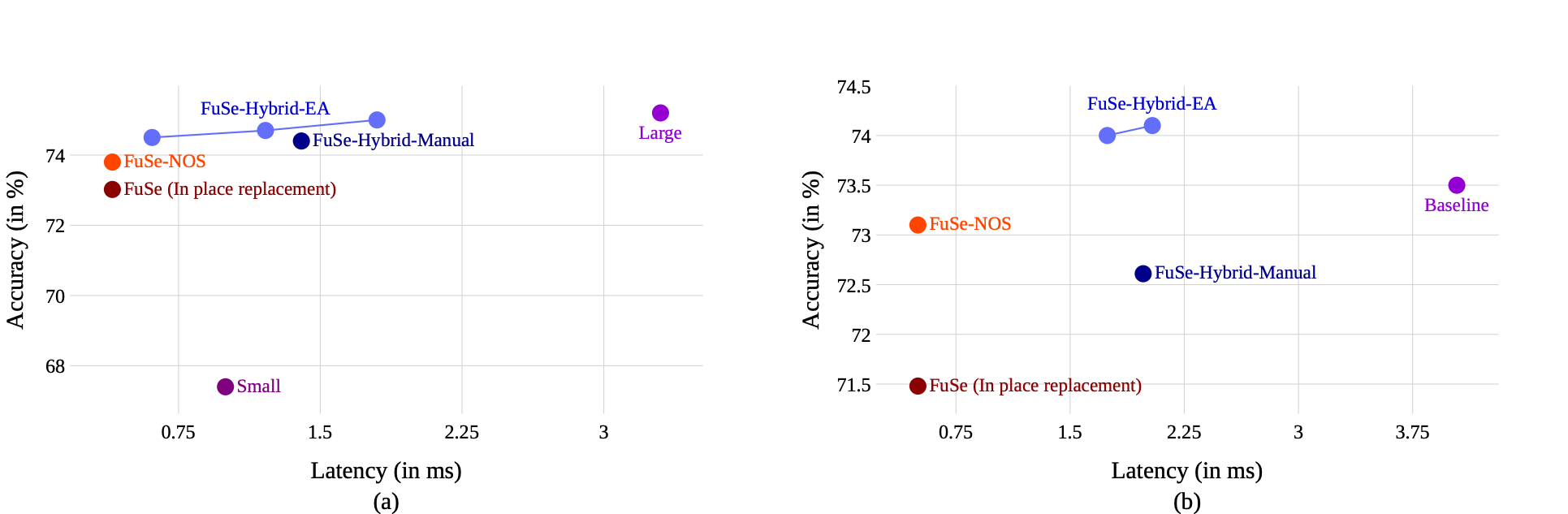}
    \caption{Accuracy vs. efficiency trade-off: (a) Hybrid MobileNetV3-Large and (b) Hybrid MnasNet-B1 networks trained using \nos~and discovered using Evolutionary Search. The pareto optimal frontier for these hybrid networks are superior relative to \fuseconv~networks trained with in-place replacement and \fuseconv~networks trained with \nos~without hybrid search.}
    \label{fig:pareto}
\end{figure}

We now report results of searching hybrid networks (combining \fuseconv~and \dwsepconv) using evolutionary search.
In Figure \ref{fig:pareto}, we plot the pareto optimal frontier of these networks, for both MobileNetV3-Large and MnasNet-B1. 
We compare these hybrid networks with the ones trained via in-place replacement and with NOS without hybrid search, as discussed in the preceding sub-sections. 

For MobileNetv3-Large, the most accurate \fuseconv~network has an accuracy of 74.9\%.
In comparison, the baseline model with \dwsepconvs~ has an accuracy of 75.3\% at 1.8$\times$ the latency cost. 
Thus, by adding \nos~and EA search, we reduce the accuracy gap between \fuseconv~and \dwsepconv~from 1.5\% to 0.4\%. 
For MnasNet-B1, the accuracy of the most accurate hybrid network exceeds that of the baseline by 0.8\% while still being $2.4\times$ faster. 
This is a significant result - Our hybrid model improves on the accuracy of a network optimized by NAS to be accurate on the ImageNet data at significantly lower latency.

\begin{figure}
    \centering
    \includegraphics[width=1\textwidth]{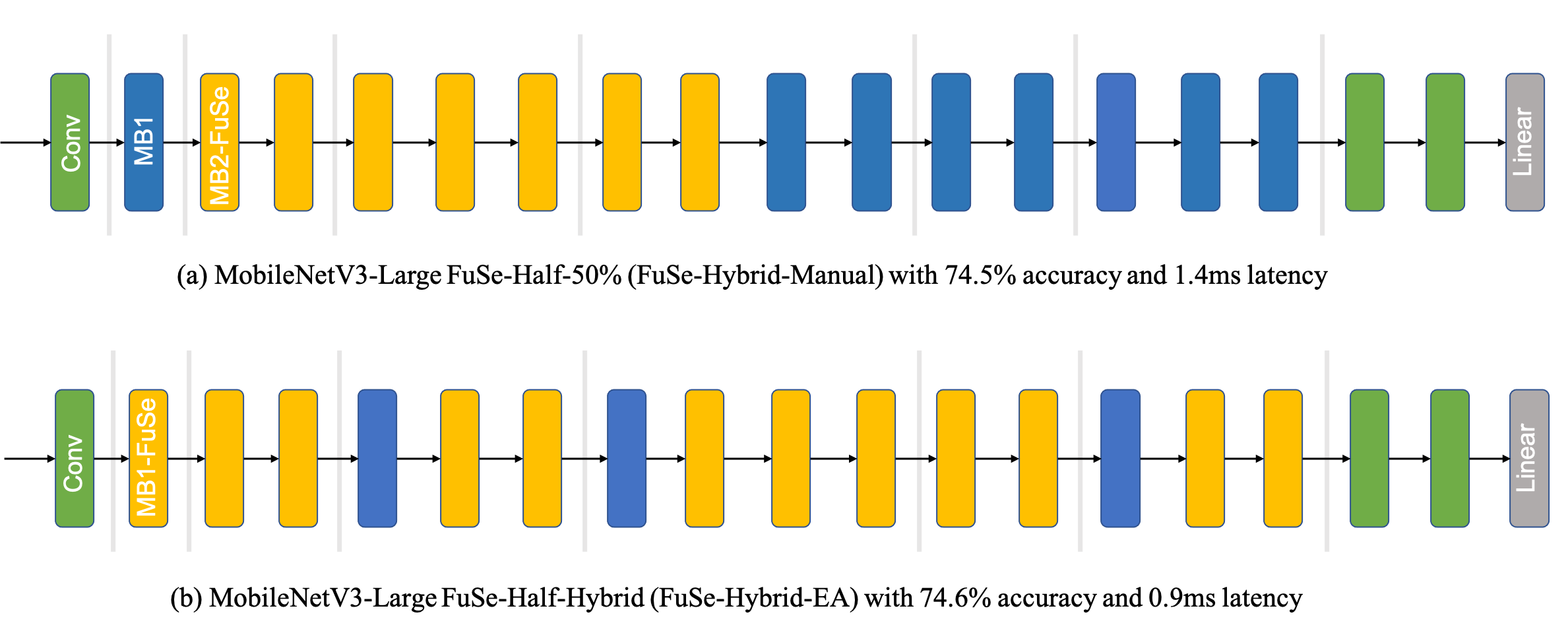}
    \caption{Visualizing MobileNetV3-Large hybrid networks: (a) Trained with in-place replacement (FuSe-Hybrid-Manual) and (b) Trained with \nos~and searched with EA (FuSe-Hybrid-EA). Clearly, \nos~coupled with EA is able to find networks that are more accurate and efficient.}
    \label{fig:nos_visualization}
\end{figure}

All the hybrid networks found using \nos~are superior to manually chosen hybrid networks that have 50\% of their \fusehalf~layers preferentially retained. 
This validates the effectiveness of the evolutionary search algorithm in identifying the trade-off between accuracy and latency.
To understand this further, we visualize in Figure \ref{fig:nos_visualization} the hybrid network discovered using EAs and compare against the manually chosen hybrid network.
We observe that the hybrid network searched with EA has more \fusehalf~layers (yellow) achieving a lower latency but still retaining the higher accuracy. 
Thus, carefully choosing which layers to convert to \fuseconv~is an important decision that is effectively optimized by EA.


\subsection{Accuracy of \fuseconv~with \nos~and Neural Architecture Search} 

\begin{figure}
    \centering
    \includegraphics[width=0.6\textwidth]{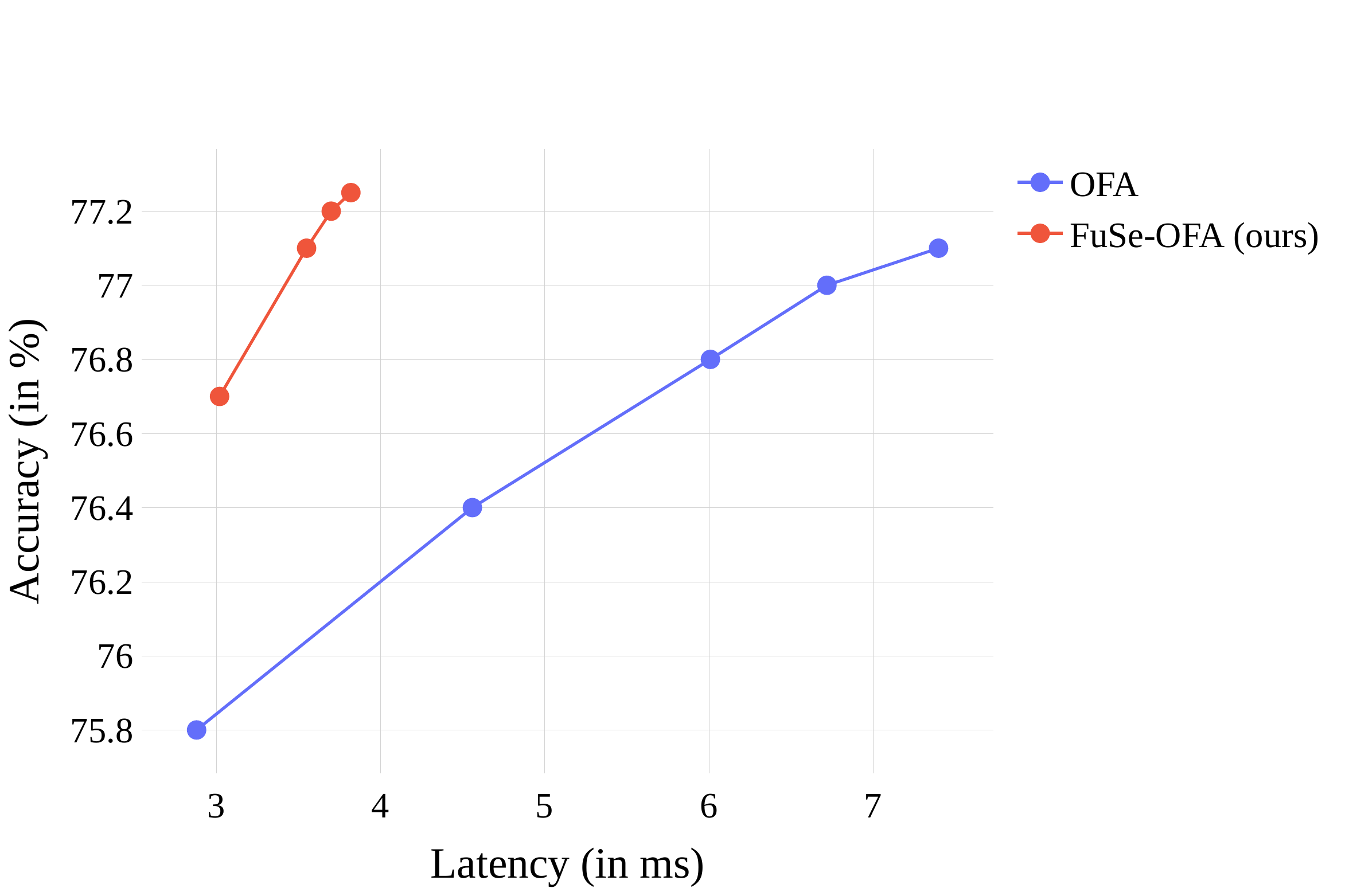}
    \caption{OFA networks with the FuSe operator consistently outperforms the baseline OFA on a 16x16 systolic-array.}
    \label{fig:ofa_pareto}
\end{figure}

We now present the results of searching for hybrid networks with Neural Architecture Search (NAS).
Specifically, we combine the search for hybrid networks into the design space of Once-For-All \cite{cai2019once} as discussed in Section~\ref{sec:nos}, which explores depth, width, and kernel size parameters.
We compare two sets of results - Models searched only with Once-For-All design space (depth, width, kernel parameters) and models searched with the choice of \fuseconv~added to the design space. 
The results are shown in \ref{fig:ofa_pareto}. 
The pareto surface of the networks searched by including the choice of \fuseconv~operators is significantly superior - it finds networks that simultaneously are more accurate and faster. 
For instance, the most accurate network that we discover with \fuseconv~has 0.1\% higher accuracy while halving the latency. 

Further, we also observe that the combining \nos~with the OFA design space helped explore the efficiency-accuracy trade-off more widely.
In particular, we achieve an accuracy of 77.2\% with \nos~and OFA, which is higher than training without \nos, training with \nos, training with \nos~and evolutionary search.
This suggests that \nos~combines effectively with standard architectural choices.

We compare the results obtained with other state-of-the-art mobile-efficient networks designed with NAS in Table \ref{table:nas_networks}.
We observe two key results: 
\begin{itemize}
    \item We find the most accurate model (FuSe-OFA-2) with an accuracy of 77.2\% matching EfficientNet-EdgeTPU-S \cite{gupta2020accelerator} which has more than 5$\times$ the MACs.
    Interestingly, the EfficientNet-EdgeTPU-S model represents an alternative to increase the utilization on systolic arrays, by using spatial convolutions. 
    Still, this model is 40\% slower than our reported model.
    Thus our model, FuSe-OFA-2 represents the state-of-the-art in efficient models for ImageNet dataset.
    \item If we prioritize latency, then our MobileNet-V3-Large FuSe-Half-Hybrid model achieves an accuracy of 74.9\% with a latency under 1ms.
    This is a significant improvement on existing networks.
\end{itemize}

\begin{table}[!h]
\centering
\vspace{0pt}
\scalebox{0.80}{
\begin{tabular}{lcccc}
\toprule
Network & \begin{tabular}[c]{@{}l@{}}ImageNet\\accuracy\end{tabular}  & \begin{tabular}[c]{@{}l@{}}MACs\\(millions)\end{tabular} & \begin{tabular}[c]{@{}l@{}}Params\\(millions)\end{tabular} & \begin{tabular}[c]{@{}l@{}}Latency on Systolic-Arrays (in ms) (batch size 1)\end{tabular} \\ 
\toprule
MnasNet-B1\cite{tan2019mnasnet}                      &  73.5  & 325  & 4.38 &  4.04 \\
MnasNet-B1 FuSe-Half (ours)                                &  73.4   & 305  & 4.25 & 0.50 \\
MnasNet-B1 FuSe-Half-Hybrid (ours)                         & 73.8    & 310   & 4.30   & 1.74 \\
\midrule
MobileNet-V3-Large\cite{sandler2018mobilenetv2}      & 75.3  & 238 & 5.47  & 3.30  \\
MobileNet-V3-Large FuSe-Half (ours)                         & 73.8  & 225 & 5.40 & 0.40 \\
\textbf{MobileNet-V3-Large FuSe-Half-Hybrid (ours)}                 & \textbf{74.9}  &  230 & 5.43 & \textbf{1.80} \\
\midrule
ProxylessNAS \cite{cai2018proxylessnas}              & 74.6   & 320 & 4.08 & 4.87 \\
\midrule 
Single-Path NAS \cite{stamoulis2019single}           &  74.7   & 332  &  4.42  & 4.25 \\
\midrule 
FBNet-C \cite{wu2019fbnet}                           &  74.9  & 382 & 5.50 &  4.70 \\
\midrule
EfficientNet-Lite0 \cite{tan2019efficientnet}        & 75.1 & 407   & 4.70 & 4.82 \\ 
EfficientNet-EdgeTPU-S \cite{gupta2020accelerator}   &  77.2  &   2351  &   5.43   & 5.35 \\
\midrule
Once-For-All (OFA) \cite{cai2019once}                &  77.1  & 369    &   6.55   & 7.40 \\
\midrule 
FuSe-OFA-1 (ours)                                    &   76.7  &  376  &   6.85   & 3.02 \\
\textbf{FuSe-OFA-2  (ours)}                          &  \textbf{77.2}   &  426   &  7.29   & \textbf{3.82} \\

\hline
\toprule
\end{tabular}
}
\caption{ImageNet performance, MACs, parameters, and latency of NAS networks on a 16x16 systolic-array.}
\label{table:nas_networks}
\end{table}

\section{Related Work}
\label{sec:related}
\noindent 
There have been many prior works that proposed solutions for efficient DNN operators. 
The efforts can be broadly grouped into these categories. 

\textbf{Fast Algorithms for Convolutions.}
For the better half of the last decade, multiple efforts focused on reducing the arithmetic complexity of convolution layers. For instance, \cite{mathieu2013fast} proposed using Fast Fourier Transforms (FFT) to perform the convolution as pointwise products in the frequency domain. This offers significant computation reduction but works only for large filter sizes that are not common in DNNs. \cite{cong2014minimizing} used Strassen's algorithm to reduce the amount of convolutions and consequently the inference time of AlexNet by 47\%. \cite{lavin2016fast} proposed the use of Winograd's minimal filtering algorithm to transform convolutions thereby reducing the number of multiplications significantly for an increase in the number of additions. All these proposals work on the algorithmic nature of how convolutions are performed. In contrast, FuSeConv operates on the structure of filters and feature maps. Hence, these fast algorithms are complementary to FuSeConv and any of these methods can be applied in conjunction to potentially reap more benefits. 

\textbf{Factorized Convolutions.}
Factorization by low-rank approximation of filters has been a popular way of reducing the computational complexity of convolutions. In the context of Deep Learning, it was first suggested in \cite{rigamonti2013learning}. Essentially, it involves decomposing one or more dimensions of the filter to improve the execution time while trying to retain the accuracy. \cite{jaderberg2014speeding} proposes two schemes to approximate the filters into a linear combination of smaller basis set of separable rank-1 filters - one along the spatial axis and the other along the channel axis. \cite{lebedev2014speeding} uses CP-decomposition to decompose a convolution layer into 4 sub-layers with smaller lower-rank kernels. \cite{kim2015compression} proposed Tucker decomposition that uses higher-order SVDs to decompose a convolution tensor into another tensor multiplied by matrices along one or more modes (each mode is a dimension of a tensor: horizontal, vertical or depth slice). SqueezeNext \cite{gholami2018squeezenext} augmented the popular squeezeNet architecture and proposed the use of a two-stage squeeze module. One of the stages is composed of spatially separable convolutions where the 3x3 spatial convolution in squeezeNet is replaced with two sequential blocks of 1x3 followed by 3x1 spatial convolution filters, while the other stage consists of element-wise addition. In general, the use of low-rank convolutions reduced the total number of parameters and offered significant computational benefits. FuSeConv is also a low-rank operation to optimize convolutions. The key differences of FuSeConv are as follows: (1) Many of the prior works that relied on mathematical techniques to perform low-rank approximations suffered high accuracy loss ($>1\%$) while FuSeConv with our proposed training methodology has comparable accuracy with the baseline, and (2) The prior works do not address the specific problem of mapping inefficiency to systolic-arrays, while FuSeConv is a systolic algorithm that enables fast inference on systolic-arrays.       

\textbf{Addressing the inefficiency of depthwise separable convolutions on accelerators.} 
\cite{gupta2020accelerator} is one of the recent works from Google that addresses the problem of inefficiency between depth-wise separable convolutions and systolic-array like accelerators. They note that the popular inverted mobile bottleneck blocks, primarily used in MobileNet family of networks \cite{sandler2018mobilenetv2} and composed of a depthwise convolution layer sandwiched between two pointwise layers, do not sufficiently utilize the EdgeTPU hardware. In order to alleviate this problem, the authors introduce an operation called fused inverted bottleneck that replaces the first pointwise layer and a depthwise convolution layer with a full spatial convolution operation. They add fused inverted bottleneck to the EfficientNet search space and use NAS to find a network specific for EdgeTPU called EfficientNet-EdgeTPU. It is noteworthy that fused inverted bottleneck consumes upto 12x more MACs for certain layers just to improve the hardware utilization. In contrast, FuSeConv provides a significant improvement in utilization and speedup for similar number of MACs relative to mobile inverted bottleneck blocks.

\section{Conclusion and looking ahead} 
\label{sec:conclusion}
\noindent

Addressing the challenge of efficient inference is vital to enable pervasive deployment of DNNs on the edge. 
Massively parallel systolic-arrays and resource-efficient \dwsepconvs~are two promising approaches. The combination of these two methods, however, is inefficient. 
In this paper, we posed three questions regarding this inefficiency: (1) Why are \dwsepconvs~inefficient on systolic-arrays?, (2) Is there a HW/SW co-design to address this inefficiency?, and (3) Can we adapt model-training to better fit this new design? 
For the first question, we formally show that \dwsepconvs~are not native systolic-algorithms and lack data-reuse or channel-wise refactoring to sufficiently utilize the systolic-array hardware. 
For the second question, we proposed a HW/SW solution with a systolic drop-in replacement operator called \fuseconv~and a customized hardware dataflow called \stos~ with low VLSI overheads.
For the third question, we proposed \nos, a novel training methodology to train more accurate \fuseconv~networks. 

We evaluated the proposed solutions on a large set of efficient networks by replacing \dwsepconvs~with \fuseconv~and observed significant improvements of $4.1-9.25\times$ in inference times on a $16\times 16$ systolic-array.
The \fuseconv~networks trained with \nos~achieved $1.5-2\%$ higher accuracy than conventional training and were as accurate as their baseline \dwsepconv~networks.
We also showed that \nos~can be combined with search exploration strategies like evolutionary algorithms and Neural Architecture Search to provide networks that significantly improve the efficiency and accuracy of computer vision models on systolic-arrays with an accuracy of $77.2\%$ on ImageNet and a latency of $3.82ms$ on a $16\times 16$ systolic-array.
These obtained models define the state-of-the-art models both in terms of accuracy and efficiency on systolic arrays. 

We conclude by reflecting on our experiences in designing \fuseconv, \stos, and \nos~and provide recommendations for further research. \\
\textbf{Resource efficient operators require special ways of training.} We saw that na\"ive in-place replacement and standard training procedures lead to accuracy degradation for the parameter-efficient \fuseconv~networks. We hypothesize the need for ``efficient training strategies'' for operators on the edge where \nos~could be one potential strategy. Other similar techniques to train parameter-efficient models may be explored as an alternative to going towards larger and larger models. \\ 
\textbf{Simple dataflow optimizations help accelerate DNN operators having low-data reuse.} Massively parallel hardware architectures assume high-data reuse as a means for optimization. However, for low-resource operators, this assumption may not be true. Thus, custom yet simple dataflow optimizations such as our proposed \stos~may be required to improve the utilization and performance of such operators on the parallel hardware. This also illustrates the need for hardware-software co-design of efficient operators and custom dataflows. \\
\textbf{The need for efficient operator search in addition to NAS.} Over the last few years, there has been a lot of work on Neural Architecture Search to find optimal network \textit{architectures} satisfying various design objectives. It is noteworthy, however, to recognize that all foundational solutions for efficient DNNs on the edge were \textit{operator} based (\dwsepconvs, squeeze-and-excite \cite{gholami2018squeezenext}, etc.). Thus, we recommend a concerted focus on designing operator design spaces (similar to network design spaces \cite{radosavovic2020designing}) and using AutoML ideas to search for efficient operators within this space.

\section{Acknowledgements}
We thank Google Cloud Platform and Robert Bosch Centre for Data Science and Artificial Intelligence, IIT Madras for their help with compute resources for this project. We thank wandb \cite{wandb} for providing a very useful experiment tracking tool that helped accelerate our progress. We thank Surya Selvam for his contributions to an earlier version of this research effort \cite{selvam2021fuseconv}.    

\bibliographystyle{acm}
\bibliography{cite.bib}

\end{document}